\documentclass[epjST]{svjour}
\usepackage{graphics}
\usepackage{array,mathtools,amssymb,booktabs,multirow,diagbox,hhline}
\newcommand{\mean}[1]{\left\langle{#1}\right\rangle}
\newcommand{\condl}{\mean{\bar q q}_l}
\newcommand{\conds}{\langle \bar s s \rangle}
\newcommand{\quarkcorlmod}{\langle \bar \psi_l \psi_l (x)  \bar \psi_l \psi_l (0)\rangle}
\newcommand{\intT}{\int_0^{1/T} \! \! \! d\tau \int d^3 \vec{x}}
\newcommand{\Od}{{\cal O}}
\newcommand{\re}{\mbox{Re}\,}
\newcommand{\im}{\mbox{Im}\,}

\begin{document}
\title{Light quarks at finite temperature: chiral  restoration and the fate of  the $U(1)_A$ symmetry}
\author{A. G\'omez Nicola \inst{1}\fnmsep\thanks{\email{gomez@ucm.es}}}
\institute{Departamento de F\'{\i}sica Te\'orica and IPARCOS. Univ. Complutense. 28040 Madrid. Spain }
\abstract{
We review recent results on the role of light quark states within the QCD phase diagram.  In particular, we will discuss how the combined use of theoretical techniques such as Effective Theories, Unitarization and Ward Identities helps to shed light on several important issues regarding  chiral symmetry restoration, building bridges with recent lattice analyses.  Special attention  will be paid to the role of chiral and $U(1)_A$ partners in the interplay between those symmetries, which is crucial to properly understand  the transition pattern. Light scalar mesons at finite temperature will be shown to be responsible for the description of susceptibilities encoding chiral and $U(1)_A$ restoration properties.} 
\maketitle
\section{Introduction}
\label{sec:intro}

\begin{figure}[h]
\hspace*{-0.7cm}
\resizebox{0.62\columnwidth}{!}{
\includegraphics{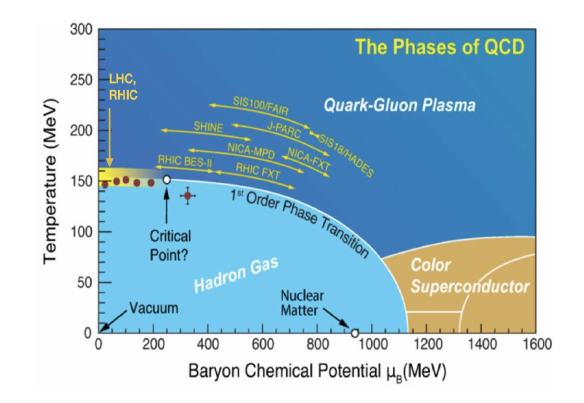}}
\hspace*{-0.5cm}
\resizebox{0.48\columnwidth}{!}{
\includegraphics{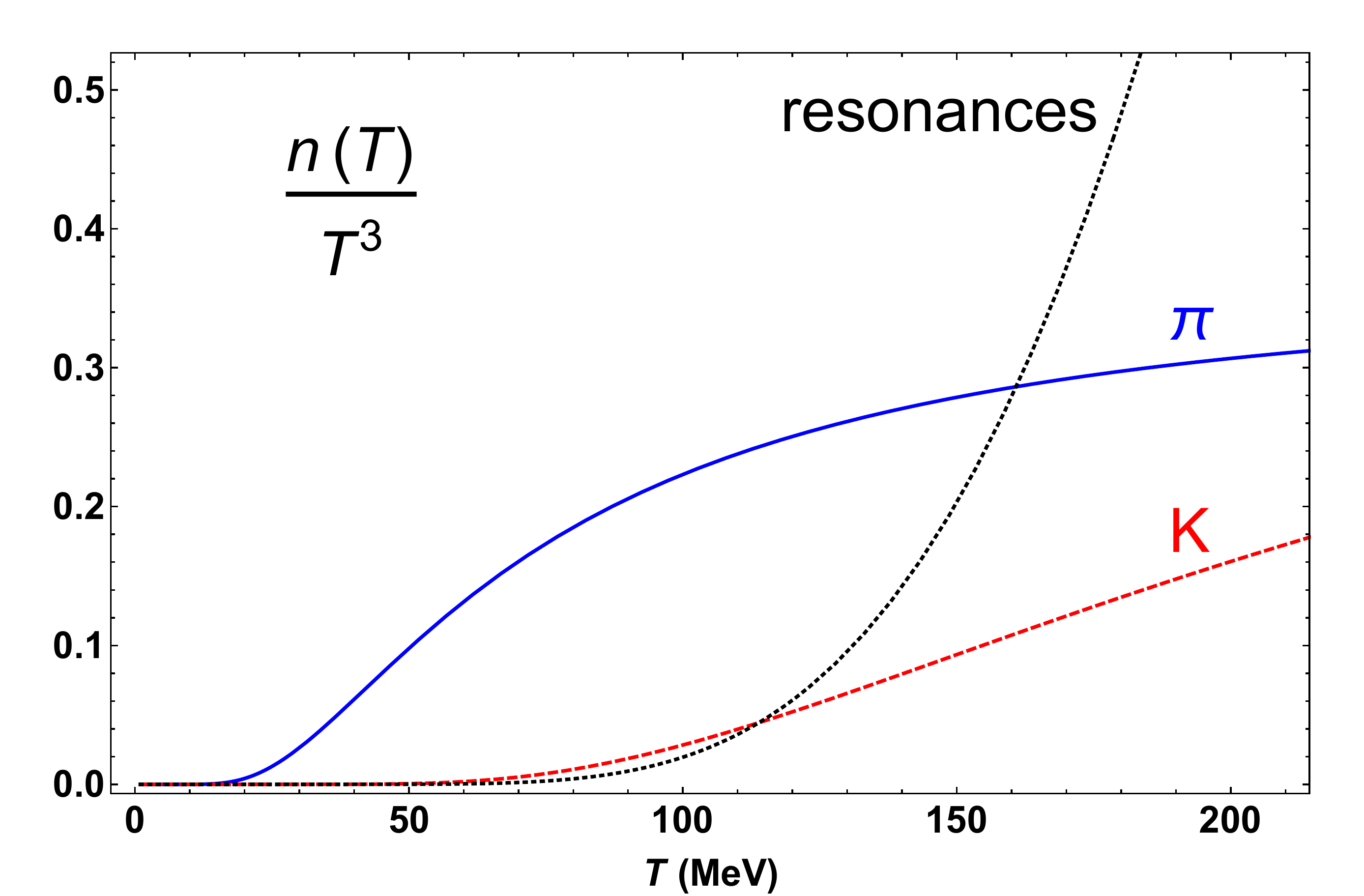}}
\begin{minipage}{\textwidth}
\caption[ex]%
{Left: Schematic view of the QCD phase diagram \cite{Bazavov:2019lgz}. Right: Free particle density for pions, kaons and the rest of resonant hadron states below 2 GeV according to the HRG approach in \cite{Jankowski:2012ms}.}
\end{minipage}
\label{fig:number}       
\end{figure}

Chiral Symmetry Restoration (CSR) under extreme  conditions of temperature and density is one of the most relevant topics concerning our present knowledge of the Quantum Chromodynamics (QCD) phase diagram, which is schematically depicted in Fig. \ref{fig:number} 
\footnote{Reprinted  by permission from  \cite{Bazavov:2019lgz},  Copyright 2019 by Springer Nature.}. There is  overwhelming evidence of the existence of a QCD transition where restoration of the chiral symmetry and deconfinement   take place, predicted from lattice simulations \cite{Aoki:2009sc,Borsanyi:2010bp,Bazavov:2011nk,Bazavov:2014pvz,Bazavov:2018mes,Ratti:2018ksb,Bazavov:2019lgz}  which has been recently supported   from  analyses of  experimental data  of Relativistic Heavy-Ion Collisions within the so called Beam Energy Scan (BES) program \cite{Adamczyk:2017iwn,Andronic:2017pug}. In those collisions, the system evolves from an initial Quark-Gluon-Plasma (QGP) phase to a hadron gas, crossing many interesting phases such as CSR and deconfinement,  from the initial highly nonequilibrated regime, which rapidly reaches local thermal equilibrium and passes through chemical and kinetic freeze out.   On the other hand, the region of low temperatures and high baryon density opens up the possibility of reaching  phases of dense quark matter such as Color Superconductivity  \cite{Alford:2007xm}, realizable in principle in astrophysical systems such as neutrons stars.

Through the analysis of the main observables involved, lattice and experimental collaborations have been able to explore quite deeply the phase diagram features for finite temperature $T$ and not too high baryon chemical potential $\mu_B$. Since lattice simulations are affected by the so called sign problem for $\mu_B\neq 0$, which has been circumvented by several methods \cite{DElia:2002tig,deForcrand:2006pv,Fodor:2004nz,Aarts:2008wh,Bazavov:2018mes}, the most accurate results  are available for $\mu_B=0$, corresponding to the  region of central rapidity in a Heavy-Ion Collision. That regime  already encodes the main physical features of the transition and will be the main subject of  the present review. Nevertheless, it is important to point out that  a major advance has come through the BES program, where a significant region  of the phase diagram has been explored   as the energy of the collision is varied. In fact, the phase diagram for low $\mu_B$ as predicted by the lattice within  Taylor expansion methods \cite{Bazavov:2018mes}  turns out to overlap with the surfaces of constant baryon number, electric charge and strangeness, obtained by fitting hadron statistical models of particle yields and their ratios  at chemical freeze-out \cite{Andronic:2017pug}. This is certainly a reassuring step forward towards the understanding of that region of the phase diagram, which is particularly important regarding the possible existence of a critical point separating the smooth crossover for small $\mu_B$ from a first-order phase transition  whose location and properties constitute one of the important open problems in this field (see \cite{Ratti:2018ksb,Bazavov:2019lgz} for  reviews of recent results). In addition, experimental input on fluctuations of conserved charges offers an additional source of information for the phase diagram, freeze-out conditions and the critical point \cite{Bazavov:2014xya,Luo:2017faz,Ratti:2018ksb}.  

At $\mu_B=0$, lattice simulations have established quite firmly that the transition is a smooth crossover at $T_c\simeq$ 155 MeV for physical quark masses \cite{Aoki:2009sc,Borsanyi:2010bp,Bazavov:2011nk,Bazavov:2014pvz,Bazavov:2018mes} usually identified as the peak position of the scalar susceptibility (see below) whereas other observables could lead to different $T_c$ estimates. As the light chiral limit of vanishing $m_{u,d}$ masses is reached, the transition is expected to become a true phase transition, possibly of second order \cite{Pisarski:1983ms,Smilga:1995qf} and the suppression of the explicit chiral symmetry breaking quark mass effect leads to a reduction of the transition temperature with respect to the physical case, as also seen  in lattice simulations where it drops down to  $T_c^0\simeq$ 129 MeV \cite{Ding:2019prx}. 

It is important to remark that, in addition to the critical point and other issues related to $\mu_B\neq 0$, there are important aspects regarding the chiral transition that are not fully understood and are the subject of current activity in this field, both within lattice and  theoretical approaches. In particular,  it remains to properly understand  the nature of the transition, mostly its  order and  universality class in the light chiral limit, as well as its interplay with the restoration of the $U(1)_A$ symmetry. Those two aspects are actually intimately connected, as we will discuss in section \ref{sec:nature}. One of the main purposes of the present work is actually to discuss specifically the present status of those problems.

Chiral symmetry $SU(N_f)_V\times SU(N_f)_A$, where $N_f=2,3$ is the number of light quark flavours,  is intrinsically linked to the low-energy  sector of QCD. In vacuum, that sector  is described by the chiral symmetry breaking pattern to the vector $SU_V(N_f)$, which governs the lowest energy states and excitations. Therefore, a consistent theoretical description of the evolution towards CSR from below the transition has necessarily to involve Effective Theories in the hadron sector, whose lightest states are the pseudo-Nambu Goldstone Bosons (NGB) of the chiral symmetry, i.e., pions for $N_f=2$ plus kaons and eta (the octet member) for $N_f=3$. The low-energy Effective Theory governing the dynamics of those states is Chiral Perturbation Theory (ChPT)   \cite{Weinberg:1978kz,Gasser:1983yg,Gasser:1984gg}. 

In  principle, effects due to heavier hadron states should become gradually important as  temperature increases, typically weighted  by their Boltzmann weight $e^{-M_i/T}$. That is actually the case for most relevant thermodynamical observables. As an indication of this trend, we show in Fig. \ref{fig:number} the free particle density as a function of temperature for pions, kaons and the rest of hadron species below 2 GeV entering within the so-called Hadron Resonance Gas (HRG) approach, where hadron interactions are assumed to be dominated by resonant channels and the corresponding resonances are accounted for essentially as free states in the partition function
\cite{Hagedorn:1968zz,Karsch:2003vd,Huovinen:2009yb,Jankowski:2012ms}. The HRG has actually been quite fruitful to describe the main properties of the transition. Despite that general trend, there are several  quantities  of interest for the phase diagram where the light quark sector provides already the dominant   contribution.  Examples of the latter will be specifically discussed  in sections \ref{sec:theo} and \ref{sec:nature}.

With all the above motivation in mind, the present work attempts to review recent advances on the QCD phase diagram, paying special attention to the role of the light quark sector. Thus, we will discuss in section \ref{sec:signals}   the main signals for CSR currently being explored in lattice and theoretical analyses, highlighting recent important results. In section \ref{sec:theo} we will review some of the main theoretical approaches  within effective hadron field theories, emphasizing the importance of thermal interactions and unitarity. Special attention will be paid in those two sections to the role of chiral partners in the light sector.  In section \ref{sec:nature} we will tackle  the problem of the nature of the chiral transition, discussing  its current status and its  connection  with the restoration of the $U(1)_A$ symmetry.

\section{Footprints of chiral symmetry restoration} 
\label{sec:signals}

The order parameter (in the  light chiral limit) of CSR  is the light quark condensate
\begin{equation}
\condl(T)=\langle \bar u u + \bar d d \rangle  (T) =\frac{\partial}{\partial m_l}z(T),
\label{conddef}
\end{equation} 
where $\mean{\cdot}$ denote Euclidean finite-$T$ correlators, $m_l=m_u=m_d$ is the light quark mass (in the isospin limit) and $z(T)=-\lim_{V\rightarrow\infty}(T/V)\log Z$ is the free energy density and $Z$ is the QCD partition function, whose hadron  representation would be given through the effective lagrangian governing the relevant degrees of  freedom (d.o.f). The quark condensate naturally follows  the thermal behaviour of the vacuum expectation value associated to the spontaneous breakdown of the chiral group \cite{Bochkarev:1995gi}. It measures the response of the vacuum to the symmetry breaking probed by the quark mass,  pretty much analogously to the mean magnetization in a ferromagnet as a response to the applied magnetic field. In the light chiral limit $\condl$ should vanish at the phase transition as a true order parameter whereas in the physical mass case it develops an inflection  point. 

In addition, one can  extract very relevant   information from the scalar susceptibility
\begin{equation}
\chi_S (T)=-\frac{\partial}{\partial m_l} \condl(T)=\int_T {dx \left[\quarkcorlmod-\condl^2(T)\right]},
\label{susdef}
\end{equation}
where  $\displaystyle \int_T dx\equiv \intT$ and  $\psi_l^T=(u,d)$, $\condl=\langle \bar\psi_l \psi_l\rangle$. The scalar susceptibility measures the order parameter correlations and as such is expected to diverge at $T_c$ in the light chiral limit while developing a peak in the physical case. Actually, $\chi_S$ is usually a more efficient way to determine the transition temperature in the lattice, together with the chiral partners discussed below.  It must be taken into account that in the crossover scenario, the transition temperature may differ from one observable to another \cite{Bazavov:2018mes}.  As explained in the introduction, correlators in the thermal bath are usually the most efficient way to determine the main thermodynamical  properties of the system and relate them with physically measurable quantities. 

In the lattice literature, the main efforts over many years, as long as CSR is  concerned, have concentrated in the measure of the above two quantities.  The main two groups undertaking the task have been the Wuppertal-Budapest  \cite{Aoki:2009sc,Borsanyi:2010bp}  and the HotQCD-Bielefeld-BNL  one  \cite{Bazavov:2011nk,Bazavov:2014pvz,Bazavov:2018mes,Bazavov:2014xya}.  Using different  lattice actions and configurations, both have agreed over the last few years on the main features of the  transition. Let us present some selection of  recent results of those groups regarding the quark condensate and the scalar susceptibility. 
\begin{figure}[h]
\hspace*{-0.55cm}
\resizebox{0.5\columnwidth}{!}{
\includegraphics{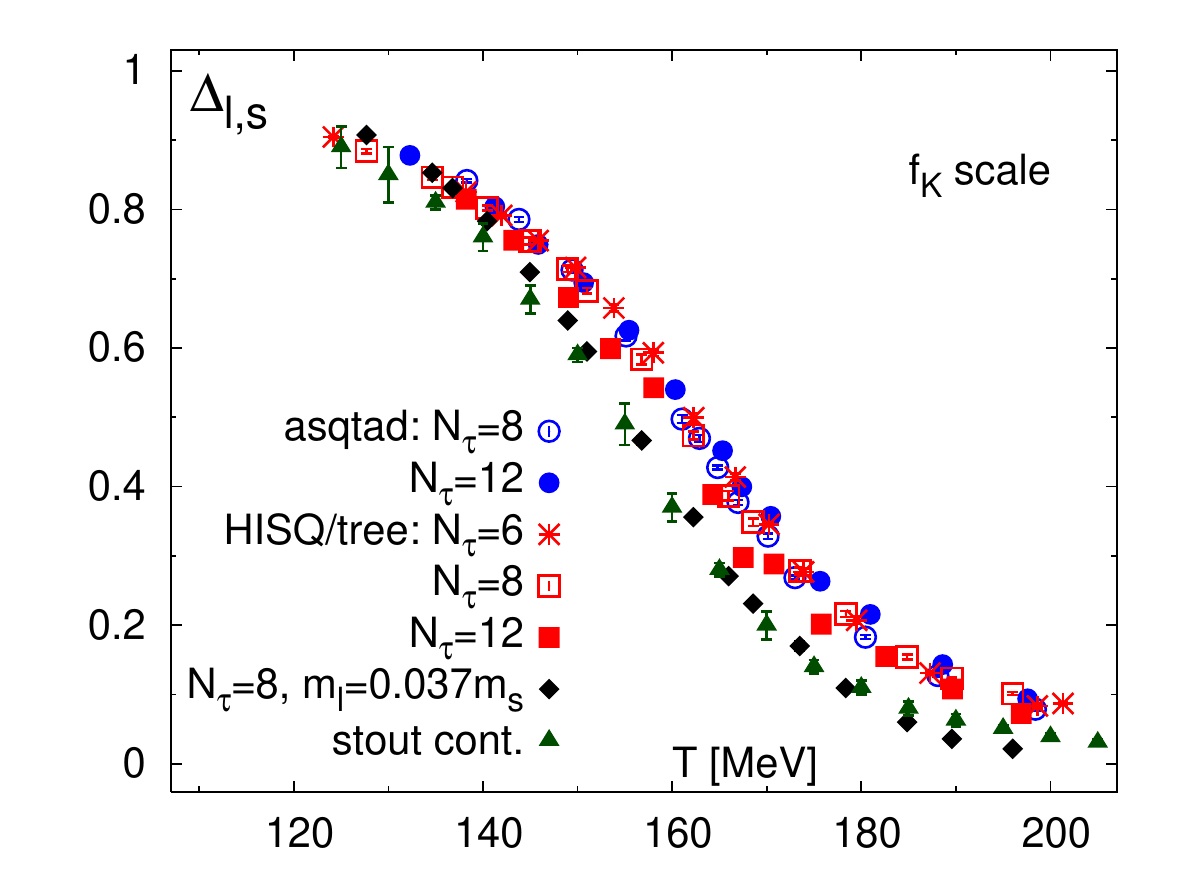}}
\hspace*{-0.5cm}
\resizebox{0.6\columnwidth}{!}{
\includegraphics{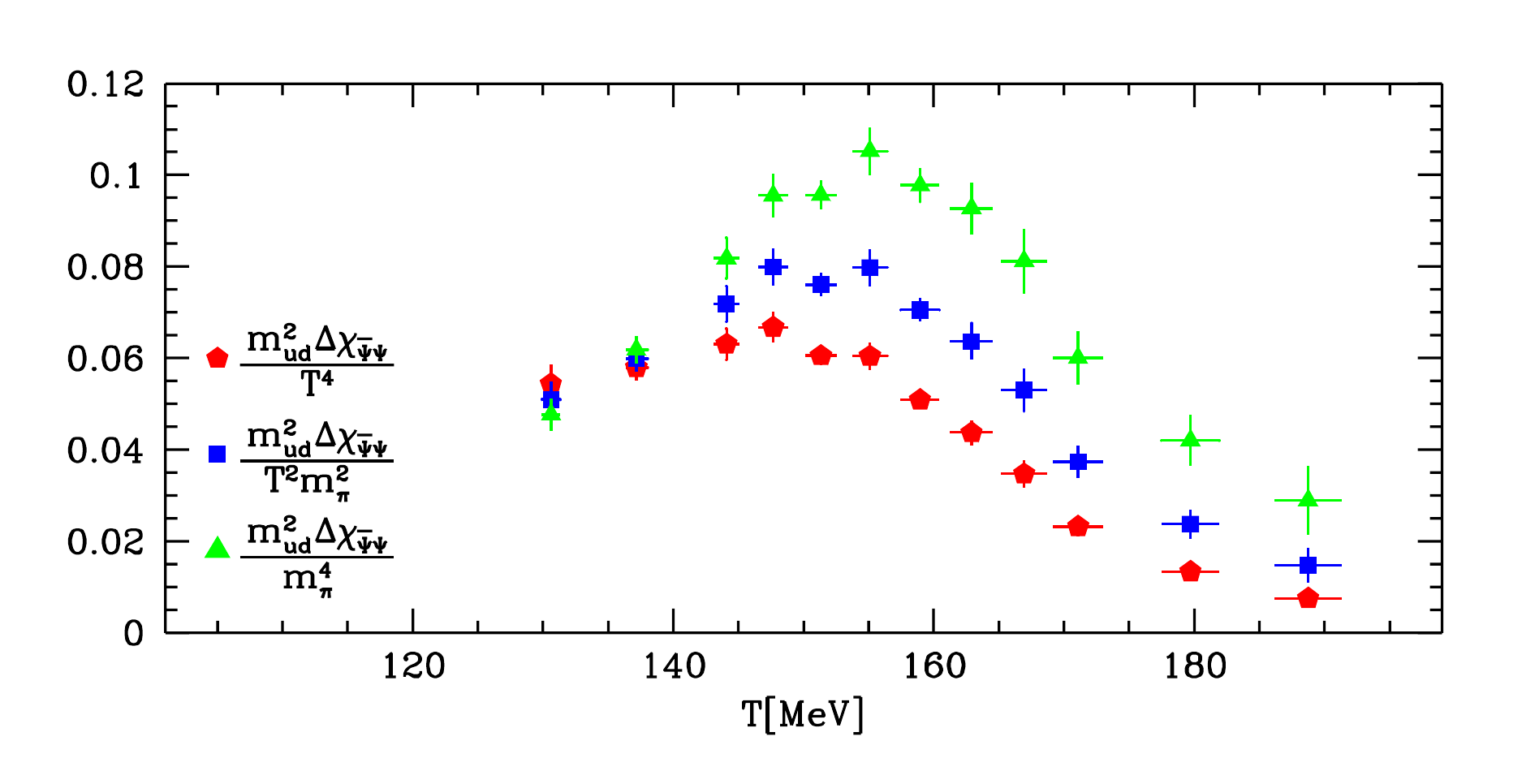}}
\caption[Caption for LOF]{Left: Subtracted  quark condensate for different actions and lattice configurations  \cite{Bazavov:2011nk}.  Right: Subtracted scalar susceptibility \cite{Aoki:2009sc}.}
\label{fig:condsuslat}       
\end{figure}

In Fig.\ref{fig:condsuslat}, we plot the subtracted quark condensate calculated in \cite{Bazavov:2011nk} \footnote{Reprinted with permission from \cite{Bazavov:2011nk} Copyright 2012 by the
American Physical Society.}. It must be taken into account that lattice quark condensates are affected by $T=0$ finite-size divergences typically scaling  as $\langle \bar q_i q_i \rangle \sim m_i/a$ with $m_i$ the  quark mass and $a$ the lattice spacing and therefore one needs to consider condensate combinations that  are  free of those divergences and still conserve the main features of the order parameter. The combination $\Delta_{l,s}=\condl-(m_l/m_s)\conds$, plotted in Fig.\ref{fig:condsuslat} normalized to  its $T=0$ value, bears that  property since the strange condensate $\conds$ decreases much slower  with $T$ than $\condl$ due to the stronger chiral symmetry breaking $m_s\gg m_l$. The inflection point around $T_c$ is clearly seen in  the figure, where results corresponding to different lattice actions (asqtad, HISQ/tree) and temporal extent $N_\tau$ are shown for the ratio $m_l/m_s=0.05$, quite  close to the physical value $m_l/m_s\simeq0.037$  \cite{Aoki:2019cca}. The continuum extrapolation is also shown. In the same figure we also plot the subtracted scalar susceptibility $\Delta\chi=\chi(T)-\chi(0)$  with different normalizations, showing the expected peak around the transition, now from the lattice collaboration  \cite{Aoki:2009sc} \footnote{Reprinted with permission from \cite{Aoki:2009sc}, Copyright 2009 by Institute of Physics Publishing.}. No significant improvement  of these data has taken place in the physical limit over the last few years. 

\begin{figure}[h]
\centering
\resizebox{0.75\columnwidth}{!}{
\includegraphics{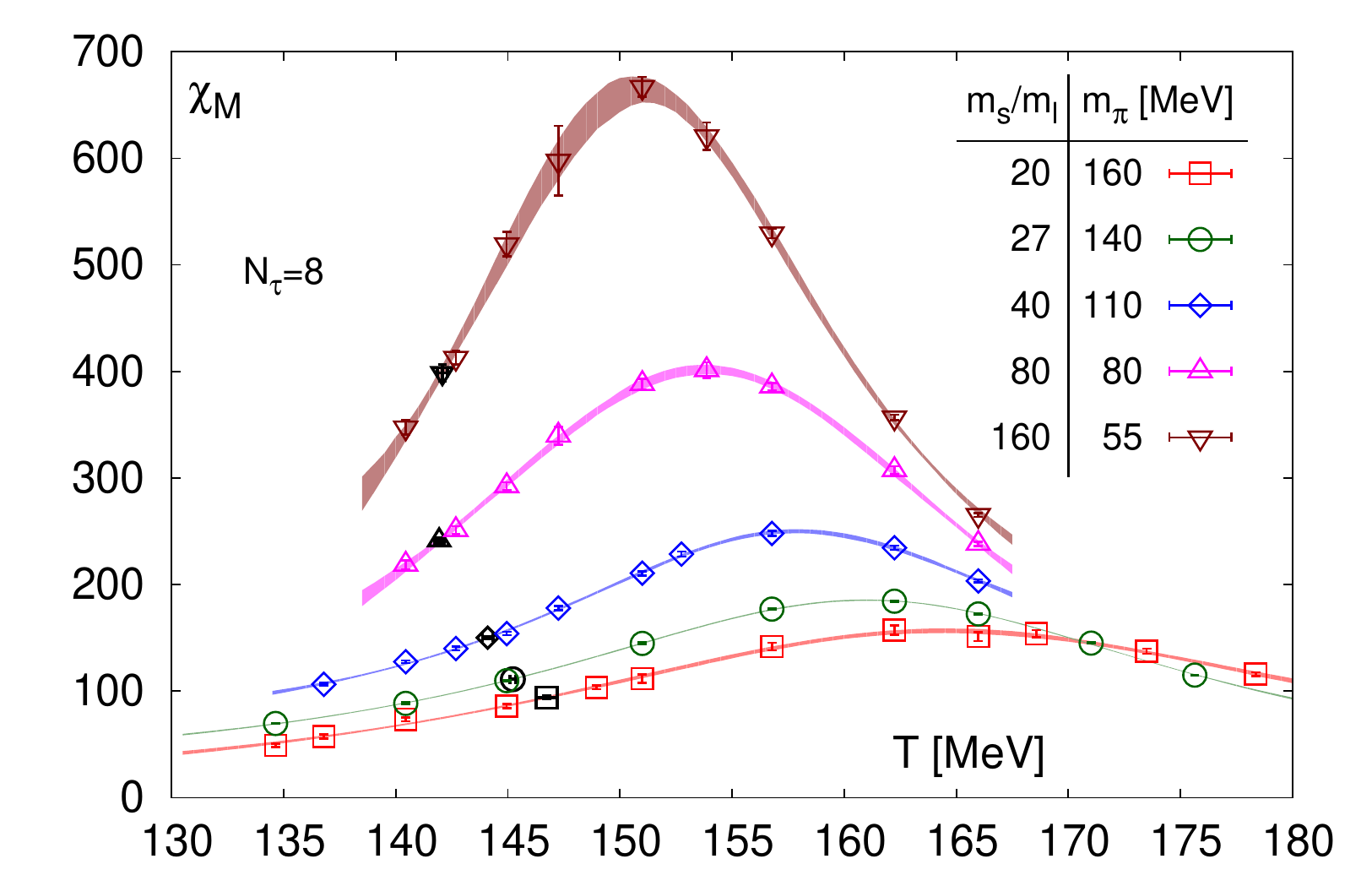}}
\caption{Evolution of the  scalar susceptibility towards the light chiral limit from \cite{Ding:2019prx}.}
\label{fig:condsuslatchlim}       
\end{figure}

Considerable progress has also been made in the lattice analysis  of the transition as the light chiral limit is approached. In Fig.\ref{fig:condsuslatchlim}  we show results from the recent analysis in  \cite{Ding:2019prx}, where $\chi_M$, defined as the light quark mass derivative of $\Delta_{l,s}$ properly normalized, follows the scalar susceptibility    trend: the transition peak is clearly enhanced and its position moves to lower temperatures as the light-strange quark mass  ratio is  reduced. The lattice setup and resolution used in \cite{Ding:2019prx} are the same as in previous works of the same group \cite{Bazavov:2011nk,Bazavov:2014pvz}. 

Let us comment now on another set of observables providing direct information about CSR, namely the so called chiral partners. The basic idea is that two quark bilinears  representing meson operators that can be connected through a $SU(N_f)_A$ rotation would become degenerate (chiral partners) if CSR takes place. Therefore, one should observe degeneration of observables constructed out of  correlators of those bilinears, like screening masses and susceptibilities.  Actually, following the same guideline, the study of  particular sets of operators  allows to analyze not only their chiral degeneration but other symmetry patterns such as $U(1)_A$ asymptotic restoration, as we will discuss in section  \ref{sec:nature}. 

Consider for instance the quark bilinears for the following isospin channels for total angular momentum $J=0$ corresponding to the pseudoscalar and scalar  meson nonets:
\begin{eqnarray}
I=0&\rightarrow& \eta_l=i\bar\psi_l\gamma_5 \psi_l, \quad \eta_s=i\bar s \gamma_5 s, \quad \sigma_l=\bar\psi_l \psi_l, \quad  \sigma_s=\bar s  s
\nonumber\\
I=1&\rightarrow& \pi^a=i\bar\psi_l\gamma_5\tau^a\psi_l, \quad \delta^a=\bar\psi_l \tau^a \psi_l \ (a=1,2,3)
\nonumber\\
I=1/2&\rightarrow&   K^a=i\bar\psi  \lambda^a  \psi, \quad   \kappa^a=i\bar\psi  \lambda^a  \psi \ (a=4,5,6,7)
\label{nonets}
\end{eqnarray}
with $\psi^T=(u,d,s)$ and $\tau^a,\lambda^a$ Pauli and Gell-Mann matrices respectively. The lowest energy states in the hadron spectrum corresponding to the quantum numbers of the above operators are, on the one hand, the light/strange components of the $\eta/\eta'$, the pion an the kaon for the pseudoscalars $\eta_l,\eta_s,\pi^a,K^a$ respectively, and on the other hand the light/strange components of the $f_0(500) \mbox{(or $\sigma$)}/f_0(980)$, the $a_0(980)$ and the $K^*(800)$ (or $\kappa$) for the scalars $\sigma_l,\sigma_s,\delta^a,\kappa^a$ respectively.

With a proper chiral $SU(2)_A$ transformations  on the quark fields one can connect the bilinears $\pi^a\,\xleftrightarrow{SU(2)_A}\sigma, \quad \delta^a\xleftrightarrow{SU(2)_A}\eta_l$, which would become chiral partners,  while a $U(1)_A$ rotation connects  
 $\pi^a\xleftrightarrow{U(1)_A} \delta^a,  \quad \sigma\xleftrightarrow{U(1)_A}\eta_l$.  In particular, under the $SU(2)_A$ transformation $\psi_l\rightarrow \exp(i\alpha_A^b\tau^b\gamma_5/2)\psi_l$ the above bilinears transform infinitesimally as
 
 \begin{align}
&\delta \pi^a (y)/\delta\alpha_A^b(x)=-\delta_{ab}\delta (x-y)\sigma_l(x),&&\delta\sigma_l(y)/\delta\alpha_A^b(x)=\delta(x-y)\pi^b(x),\nonumber\\  
&\delta \delta^a (y)/\delta\alpha_A^b(x)=\delta_{ab}\delta (x-y)\eta_l(x),&&\delta\eta_l(y)/\delta\alpha_A^b(x)=-\delta(x-y)\delta^b(x) 
\label{chitransbi}
\end{align}
with $a,b=1,2,3$,   whereas for $U(1)_A$ transformations $\psi_l\rightarrow \exp(i\alpha_A\gamma_5/2)\psi_l$ one has
\begin{align} 
&\delta\pi^{a}(y)/\delta\alpha_A(x)=-\delta(x-y)\delta^a(x),&&\delta\delta^{a}(y)/\delta\alpha_A(x)=\delta(x-y)\pi^a (x),\nonumber\\ 
&\delta\sigma_l(y)/\delta\alpha_A(x)=\delta(x-y)\eta_l (x),&&\delta\eta_l(y)/\delta\alpha_A(x)=-\delta(x-y)\sigma_l (x).
\label{diagtrans}
\end{align}

 As we will discuss in section \ref{sec:nature}, the  interplay between chiral $SU(2)_V\times SU(2)_A\approx O(4)$ and $U(1)_A$ symmetries  plays a crucial role regarding the nature of the transition. Therefore, the above $I=0,1$ set of operators and their corresponding correlators are suitable probes in that context. Actually, the  difference of susceptibilities 
  $\chi_{5,disc}=\frac{1}{4}\left[\chi_P^\pi-\chi_P^{\eta_l}\right]$ is customarily used in the lattice as  the parameter measuring $O(4)\times U(1)_A$  restoration. As for the $I=1/2$ sector, the 
 $K-\kappa$ bilinears would degenerate both with a $SU(2)_A$ and a $U(1)_A$ rotation, which, as we will see in section \ref{sec:nature}, offers additional interesting  possibilities for studying the interplay between chiral and $U(1)_A$ restoration, regarding in particular the role of strangeness.

\begin{figure}[h]
\resizebox{0.5\columnwidth}{!}{
\includegraphics{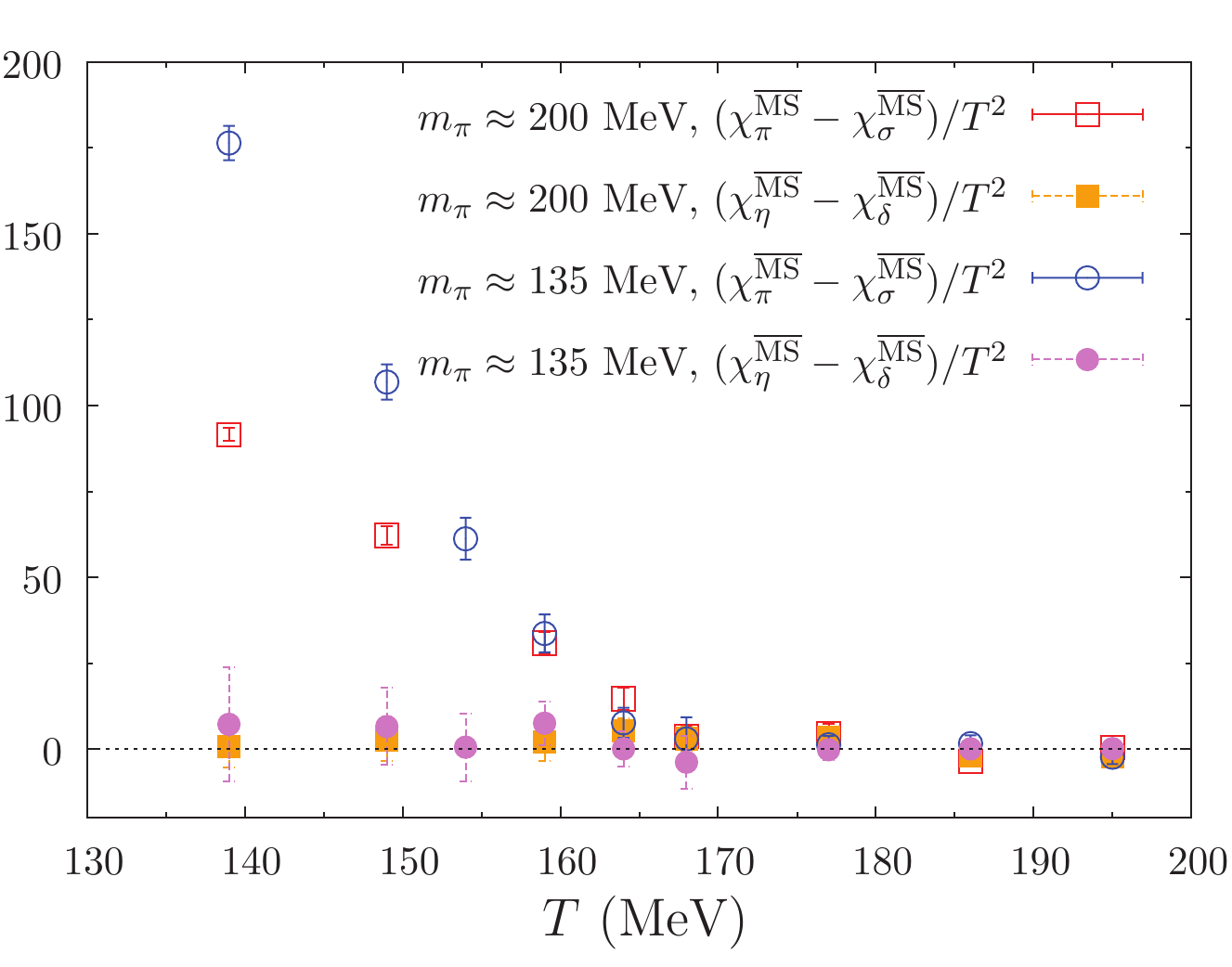}}
\resizebox{0.5\columnwidth}{!}{
\includegraphics{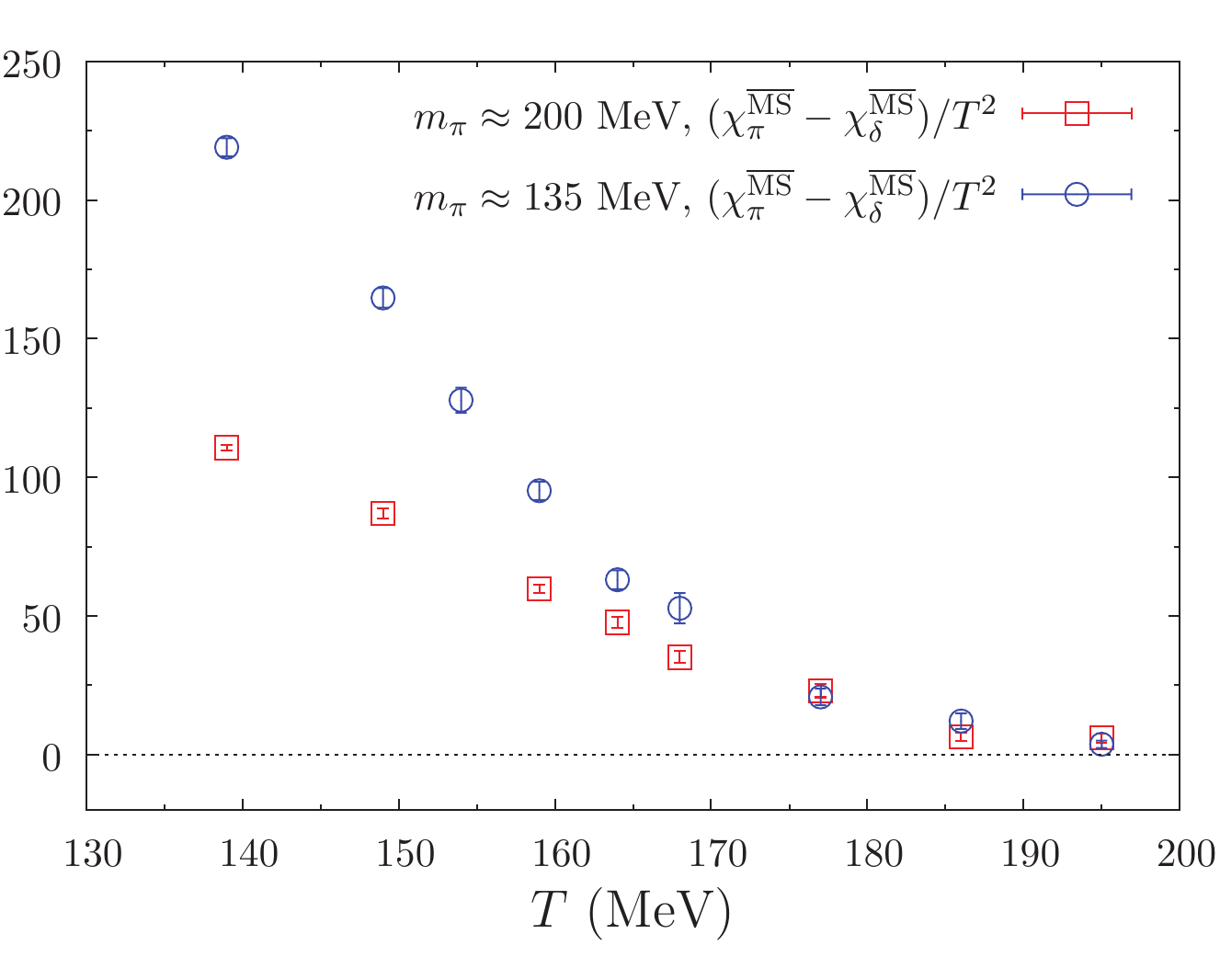}}
\caption{Results from \cite{Bhattacharya:2014ara} for chiral- (left) and $U(1)_A$- (right) partners susceptibilities.} 
\label{fig:partnerslatt21}       
\end{figure}

Chiral partners have been  explored in  lattice collaborations  over recent years for different channels, mostly through susceptibilities, defined by the integral  of the correlator over euclidean space-time at finite $T$, analogously to (\ref{susdef}), and screening masses measuring the inverse of the screening length in the exponential falloff with distance of  euclidean correlators.   We select here some recent relevant results. In Fig.\ref{fig:partnerslatt21} 
\footnote{Reprinted with permission from  \cite{Bhattacharya:2014ara} Copyright 2014 by the
American Physical Society.}
we show the results obtained by the HotQCD collaboration  \cite{Bhattacharya:2014ara,Buchoff:2013nra} for $N_f=2+1$ flavours for the different susceptibilities involved and for physical quark masses. On the one hand, CSR degeneracy of the $\pi-\sigma$ partners is clearly seen, although at a slightly higher temperature than the $T_c\simeq$ 155 MeV value corresponding to the peak of the scalar susceptibility. As commented in the Introduction, this is a consequence of the crossover nature of the transition. On the other hand,  the $\delta-\eta$ one  is subject to much larger relative uncertaintites. We also show the results of that collaboration for $U(1)_A$ partners, which under those  conditions  degenerate at larger temperatures. 

\begin{figure}[h]
\resizebox{0.5\columnwidth}{!}{
\includegraphics{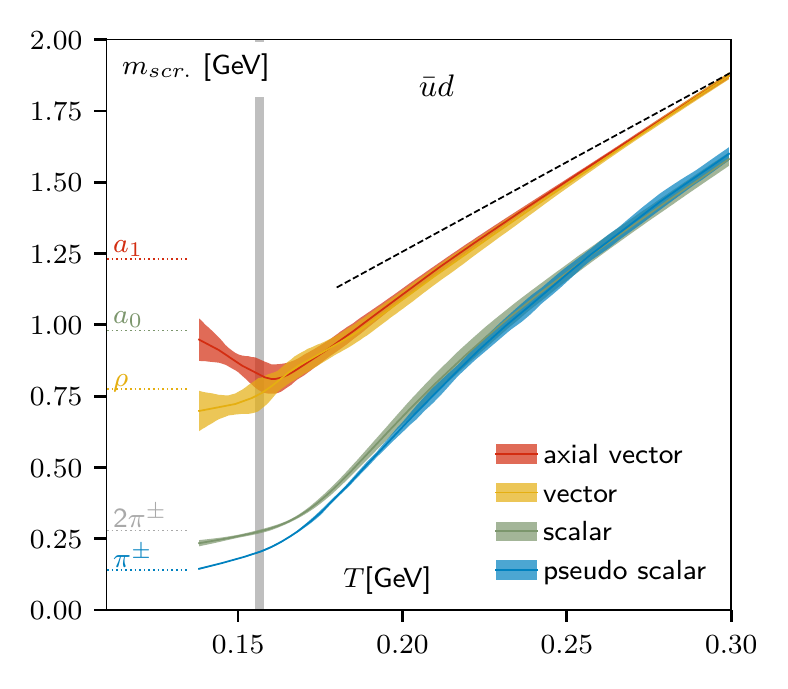}}
\resizebox{0.5\columnwidth}{!}{
\includegraphics{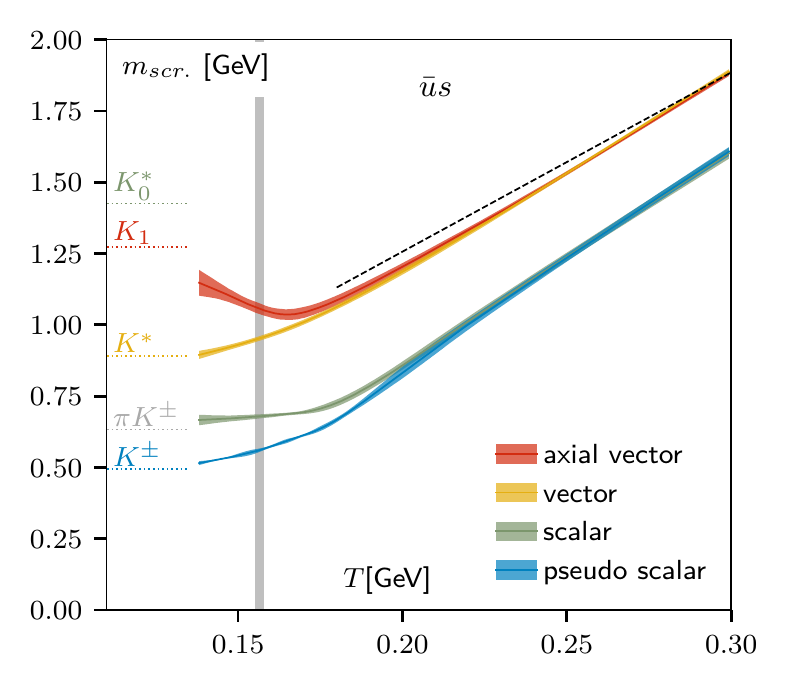}}
\caption{Results from \cite{Bazavov:2019www} for lattice screening masses (continuum extrapolation).} 
\label{fig:scrmas}       
\end{figure}

\begin{figure}[h]
\centering
\resizebox{0.75\columnwidth}{!}{
\includegraphics{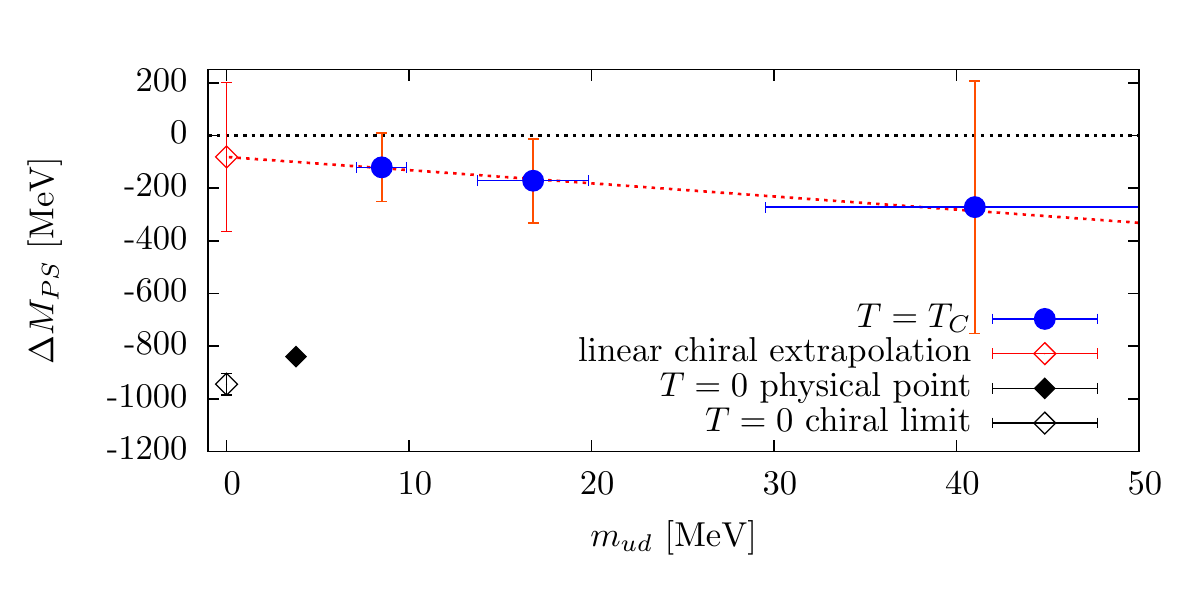}}
\caption{Results from \cite{Brandt:2016daq}  for the $\pi-\delta$ screening mass difference.} 
\label{fig:partnersBrandt}       
\end{figure}

The vector and axial-vector channels $I=J=1$, corresponding to the $\rho(770)$ and $a_1(1260)$ as lowest mass states, show also degeneracy at CSR, as it can be clearly seen for instance in the screening masses analysis performed by the Bielefeld-HotQCD group in  \cite{Cheng:2010fe} and more recently in \cite{Bazavov:2019www}. These works contain also results for other channels with partners of interest, such as the $K-\kappa$ sector corresponding to $I=1/2,J=0$, whose role we will discuss in section \ref{sec:nature}.  Selected results from   \cite{Bazavov:2019www} are  shown in Fig.\ref{fig:scrmas}.  Screening  masses for chiral partners have also been analyzed for $N_f=2$ in \cite{Brandt:2016daq}, where, apart from the $I=J=1$ sector,  masses for the scalar-pseudoscalar $I=1$ channel are also analyzed, showing that $U(1)_A$ restoration might be actually effective at the CSR temperature as the chiral  limit is approached. This is shown in Fig.\ref{fig:partnersBrandt}. Preliminary analysis by that group with larger volumes are still showing that trend but with larger $U(1)_A$ breaking  \cite{Brandt:2019ksy}. These results regarding $U(1)_A$ restoration at $T_c$ are in accordance with those analyzed for $N_f=2$ in \cite{Cossu:2013uua}, where the  correlators of the four  $I=0,1$ partners discussed above are shown to degenerate at $T_c$ at the light chiral limit, and in \cite{Tomiya:2016jwr} where  $\chi_P^\pi-\chi_S^\delta$  vanishes at $T_c$ also for $N_f=2$ in the light chiral limit.   In section \ref{sec:nature} we will discuss about this issue in more detail. 

\section{Theoretical tools}
\label{sec:theo}

Ever since the early proposals of a chiral restoration phase transition, many theoretical analyses have been developed to explore its main features.  In this sense, the range of masses, temperatures and energies involved calls for  Effective Theories and models for which hadron fields and states  constitute the main d.o.f \cite{Nicola:2020iyl}.  As explained above, higher temperatures usually  implies that heavier states have to be taken into account, although a description solely based  on the lightest modes captures the main Physics for many  relevant situations, as we will see below.

First attempts in that direction came from the $O(4)$  or Linear Sigma Model (LSM) where the d.o.f are pions  and  the explicit $\sigma$ field forming a generic $O(4)$ field with the quantum numbers corresponding  to the ($\pi^a,\sigma_l$) bilinears in \eqref{nonets}. The interaction  lagrangian can be chosen to develop spontaneous symmetry breaking $O(4)\approx SU(2)\times SU(2)\rightarrow O(3)\approx SU(2)$, whose restoration can be examined  in terms of the  finite-temperature  dependence of the relevant quantities, such as the  effective potential or the in-medium spectral  modifications of the $O(4)$ field.  It was actually within that approach that the first discussions about the transition,  including its order and the connection with $U(1)_A$ restoration, were carried out \cite{Pisarski:1983ms}.  In \cite{Bochkarev:1995gi}, a detailed analysis of CSR was carried out within the LSM, comparing it with the Non-linear Sigma Model  approach, which is nothing but the leading order in the ChPT framework discussed below. In particular,  the role of  $\langle \sigma \rangle (T)$   as order parameter  compared to $\condl (T)$, as well as the interpretation of the pion decay constant $f_\pi(T)$ in that context, are examined in that work.   A detailed analysis within the LSM including the evolution of the pion and sigma self-energies with temperature can be found in  \cite{Ayala:2000px} while in the recent work \cite{Ferreres-Sole:2018djq} an analysis of the scalar susceptibility and its connection with the $\sigma$ self-energy is provided within that model.

The Nambu-Jona-Lasinio (NJL) model has also been  extensively used  to analyze properties related to CSR. Thus, the degeneracy of  $\pi-\sigma$ chiral partners at finite  temperature  and baryon density,  through the analysis of their respective self-energies within the NJL, has been studied in \cite{Hatsuda:1985eb,Bernard:1987im}.  Coupling the NJL model to the Polyakov loop (PNJL model), the order parameter for deconfinement, has also recently allowed  to study  degeneration of the chiral  and $U(1)_A$ partners \cite{Costa:2008dp} as well as many more properties of the QCD phase diagram including the critical point \cite{Costa:2010zw}. 

A more rigorous approach to the meson gas and its properties has been  provided by the ChPT scheme. The main ideas of the ChPT approach rely on the construction of the most general effective lagrangian to every order in the generic energy expansion in $p^n$ powers (meson derivatives, momenta, masses and temperature) together with a consistent chiral power counting for loop contributions and the determination the corresponding Low Energy Constants (LEC)  associated to every term \cite{Weinberg:1978kz,Gasser:1983yg,Gasser:1984gg}.  Thus, the lightest mesons $\pi$, $K$, $\eta$ are described within this scheme as pseudo NGB  of  chiral symmetry breaking and the same formalism can be extended to heavier states such as vector mesons or nucleons \cite{Pich:1995bw}. Likewise, the $\eta'$ can be incorporated adding the  $1/N_c$ counting since the $\eta'$ mass comes from the chiral anomaly  and is therefore suppressed for large $N_c$, giving rise to the so-called U(3) ChPT scheme \cite{HerreraSiklody:1996pm,Kaiser:2000gs}.  The main advantages of the ChPT framework are its consistency and model-independency, 
thus avoiding some of the usual difficulties of model descriptions such as  the large coupling of the LSM needed to reproduce the observed $f_0(500)$ $I=J=0$ pole compatible with scattering data \cite{Pelaez:2015qba}. Another benefit  of the ChPT scheme  is that one has analytic control over the light quark masses, since the theory is built upon the massless theory with just chiral symmetry requirements. This is particularly useful for the analysis of CSR where the light chiral limit  is determinant, as described in section \ref{sec:signals} above.   The main ChPT  limitation  is that the temperature applicability range lies well below the transition, although for  certain observables a description in terms of the lightest d.o.f can be fruitful enough when combined with additional tools such as Ward Identities (WIs) and Unitarity, as we will describe here. 

\begin{figure}[h]
\resizebox{0.5\columnwidth}{!}{
\includegraphics{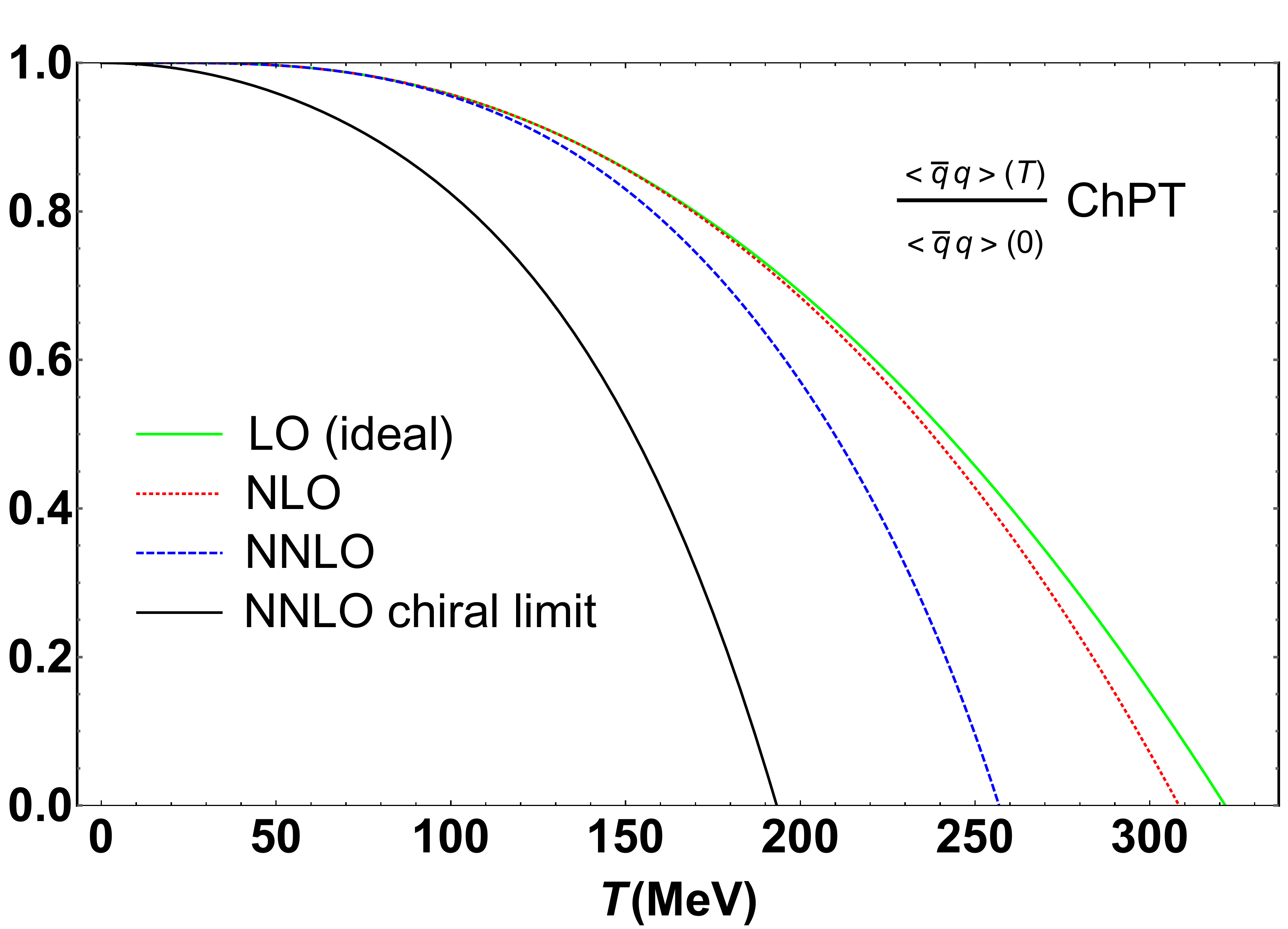}}
\resizebox{0.53\columnwidth}{!}{
\includegraphics{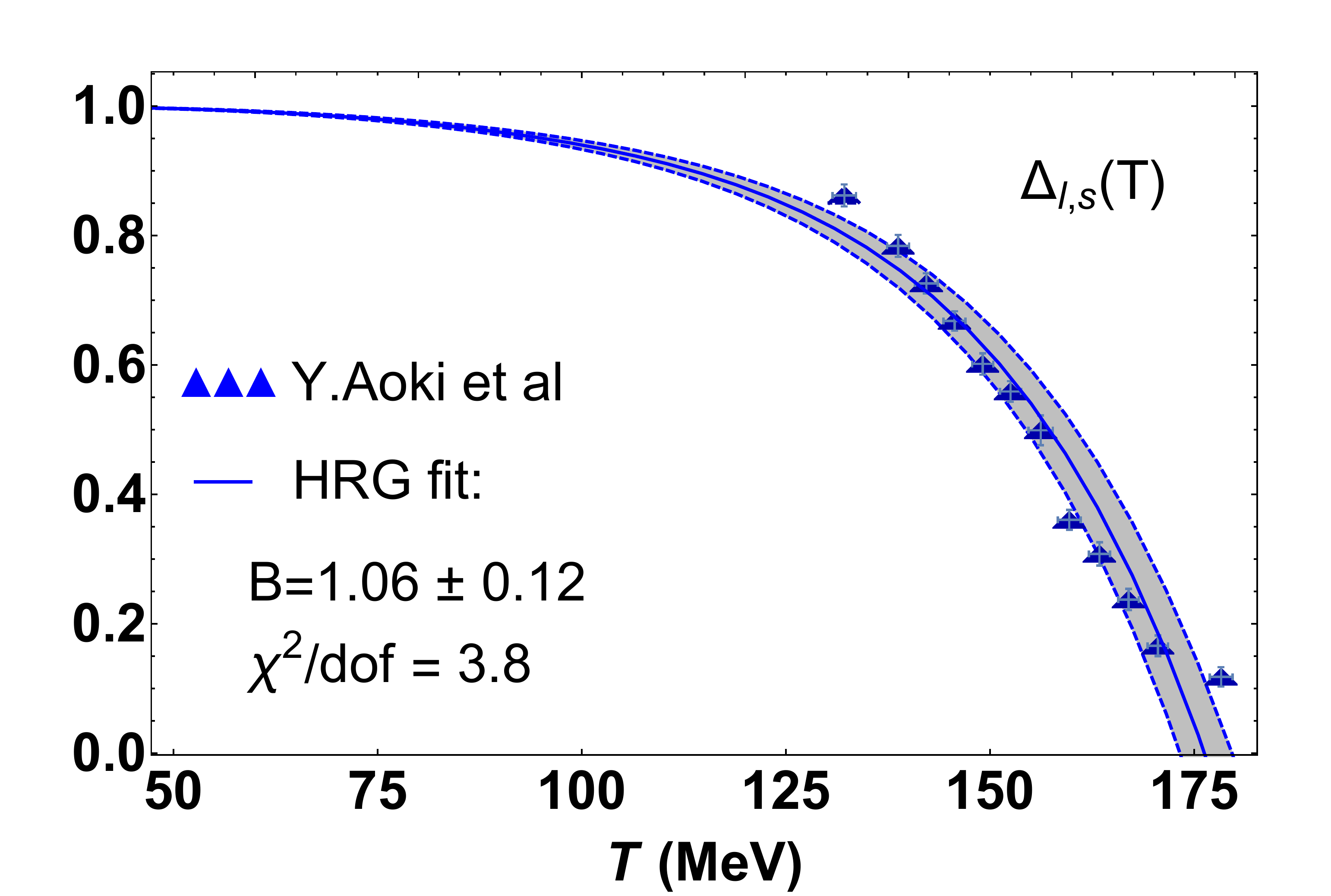}}
\caption{Left: Light quark condensate to various orders in the ChPT framework \cite{Gerber:1988tt}. Right: HRG fit to the subtracted condensate from the analysis in \cite{Jankowski:2012ms} and the lattice data in \cite{Aoki:2009sc}. The uncertaintiy bands correspond to the 95\% confidence level of the fit.} 
\label{fig:chptandhrg}       
\end{figure}

Thus,  within the ChPT framework, various finite-temperature analyses concerning the meson gas have been developed, confirming in  particular the main trends expected from CSR. Thus, early works studied the temperature dependence of the quark condensate  to lowest order \cite{Gasser:1987ah} showing the expected decrease, which was extended up to $\Od(T^8)$  (NNLO) in a thorough analysis in \cite{Gerber:1988tt}. One of the main conclusions of the latter work  is that interactions among the meson gas components in the thermal bath, which show up essentially at NNLO, are quite relevant, reducing in particular the quark condensate to values closest to the lattice expectations, while  the LO analysis corresponds to an ideal gas and the NLO corrections can be absorbed in the  renormalization of the LO.  In addition,  as mentioned above, meson thermodynamics can be consistently studied in the chiral limit in that framework, showing in particular the expected reduction of the quark condensate and hence of the extrapolated transition temperature. In Fig.\ref{fig:chptandhrg}, we show the results for the light quark condensate at different ChPT orders. In order to calibrate the effects of higher mass states, we also show the result of a fit to lattice data of the subtracted quark condensate $\Delta_{l,s}$ using the HRG approach in \cite{Jankowski:2012ms} with resonant states of masses up to 2 GeV, where the quark mass dependence of the light mesons ($\pi$,$K$,$\eta$) is taken from the ChPT predictions while for the higher mass states, the PNJL model in \cite{Blaschke:2011yv}  is used. We have allowed for an overall normalization of the free energy $z\rightarrow Bz$, fitting $B$ to lattice data below the transition, in order to account in a simple manner for the inherent HRG uncertainties such as the light mass dependence or the number of resonant states considered. The result in that figure shows that the value of $B$ obtained in the fit is compatible with unity, revealing the robustness of the HRG approach below $T_c$. However, an important observation to be taken into account is that the HRG usually predicts monotonically changing functions, as it does the ChPT approach, and hence the HRG curve for quark condensate does not have the inflection-point character around $T_c$. Nevertheless, it clearly shows that for this particular observable the effect of higher states is important to achieve a description compatible with lattice data. Other relevant results for the meson gas at finite temperature include  corrections to the  NGB dispersion relation where   a thermal dispersive contribution  defines their mean  free path \cite{Gasser:1986vb,Goity:1989gs,Schenk:1993ru,Pisarski:1996mt}, analyses  within the  low-density virial expansion which  rely on scattering amplitudes and agree with perturbative ChPT  \cite{Gerber:1988tt,Venugopalan:1992hy,Dobado:1998tv,Pelaez:2002xf,GomezNicola:2012uc,Broniowski:2015oha}, as well as the calculation of transport coefficients and  phenomenological related effects  \cite{FernandezFraile:2009mi}. 

It is also worth mentioning that the $T$-dependence of the pion decay constant within ChPT  reveals that at NLO \cite{Gasser:1986vb} it holds 

\begin{equation}
\frac{F_\pi^2(T)M_\pi^2(T)}{\condl (T)}=\frac{F_\pi^2(0)M_\pi^2(0)}{\condl (0)}\neq -m_l
\end{equation}
i.e, the Gell-Mann-Oakes-Renner (GOR) relation is broken at that order only by $T=0$ terms and since the $M_\pi (T)$ dependence is softer, $F_\pi(T)$ still follows a decreasing trend linked to that of the quark condensate, even though $F_\pi$ itself cannot be considered an order parameter \cite{Bochkarev:1995gi}. A similar trend was found in  models with explicit resonance fields \cite{Harada:1996pg}.  At NNLO the situation qualitatively changes: the space and time components of the axial current give rise to two different $F_\pi^{s,t}(T)$, which in addition develop an imaginary part. The ratio of their real parts is directly related to the velocity of pions in the heat bath, whereas their imaginary parts arise from the pion thermal width \cite{Pisarski:1996mt,Toublan:1997rr}.  GOR holds to NNLO only in the chiral limit, including temperature effects \cite{Toublan:1997rr}.

\begin{figure}[h]
\centering
\resizebox{0.75\columnwidth}{!}{
\includegraphics{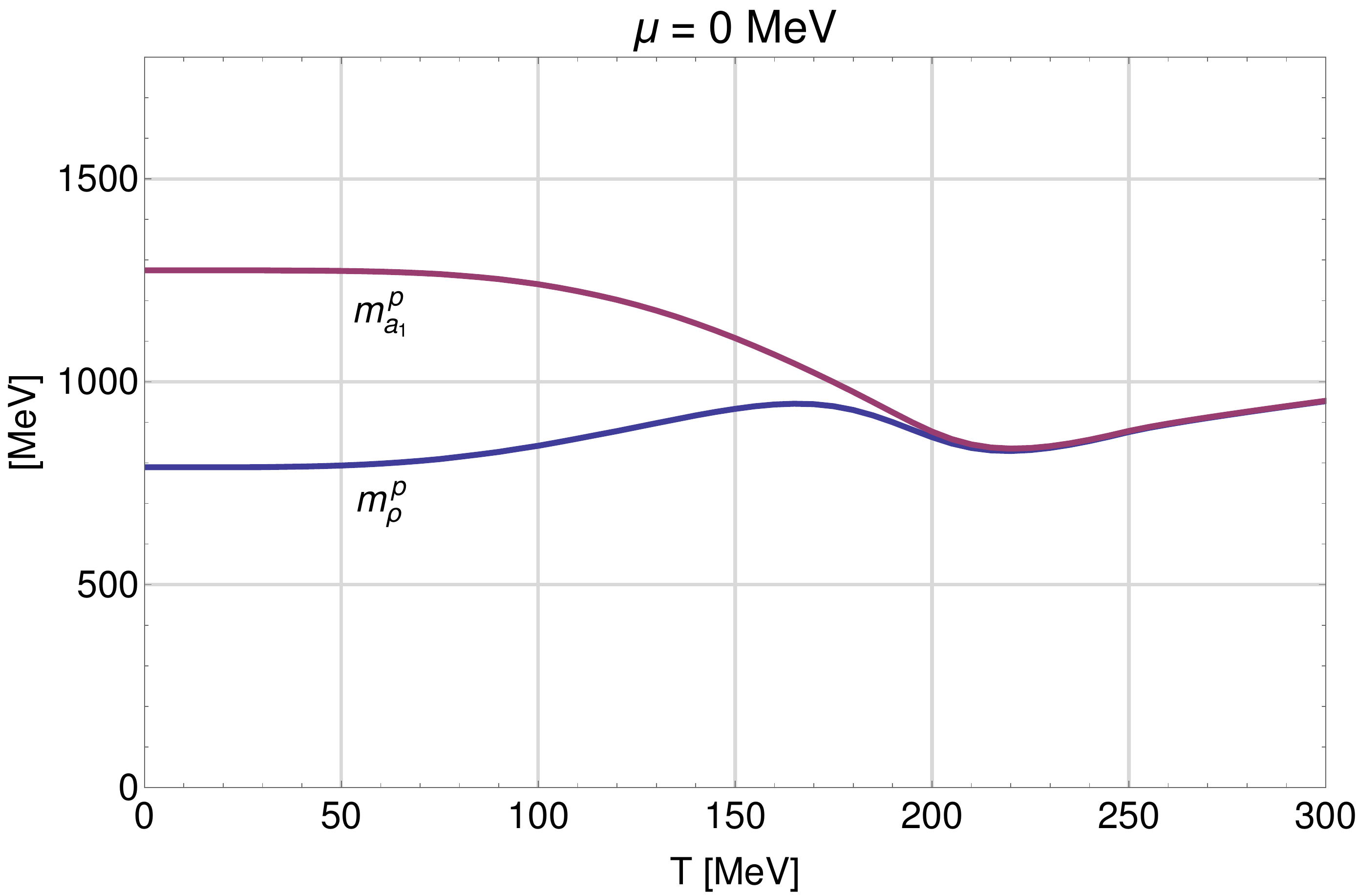}}
\caption{Modification  of  the $\rho$ and $a_1$ pole masses from \cite{Jung:2016yxl}.}
\label{fig:rhoa1masses}       
\end{figure}

Unitarization has been shown to become an important ingredient to improve the ChPT predictions in hadron phenomenology \cite{Dobado:1996ps,Oller:1997ti,GomezNicola:2001as,Pelaez:2015qba} and in particular to dynamically generate  the lightest resonant states such as   $f_0(500)$, $\rho(770)$, $a_0(980)$, $K_0^*(700)$ and  so on, with resonance parameters consistent with the expectations from the Particle Data Group  \cite{Zyla:2020zbs}. The extension of that program to finite temperature has proven to be quite useful. Thus, finite temperature pion scattering in ChPT \cite{GomezNicola:2002tn}  and its unitarization \cite{Dobado:2002xf} generate the thermal modification of the $\rho,\sigma$ resonances spectral properties, through their poles in the second Riemann sheet parametrized as $s_{pole}(T)=\left[M_p(T)-i\Gamma_p(T)/2\right]^2$. Thus, for the $\rho(770)$,   which is  a narrow resonance so that $M_p$ and $\Gamma_p$ roughly correspond  to  its mass and width, a significant broadening of $\Gamma_p(T)$ is obtained, compatible with the expectations from  other models and from the dilepton excess around the $\rho$ region  observed in heavy-ion collisions \cite{Rapp:2014hha}.  Actually, the in-medium modifications of the spectral properties of vector and axial-vector mesons and their theoretical and phenomenological implications has been the subject of very intensive work over recent years within different models describing those meson  states explicitly in the lagrangian, such as Vector Meson Dominance, gauged LSM or Hidden Local Symmetry \cite{Rapp:1999ej,Jung:2016yxl}. Particularly important for the issues analyzed in the present review is the analysis of the degeneration of the chiral partners $\rho$-$a_1$ and therefore of their spectral functions. As a highlight of those analysis,  we show  in Fig.\ref{fig:rhoa1masses}  
\footnote{Reprinted with permission from \cite{Jung:2016yxl} Copyright 2017 by the
American Physical Society.}
the temperature  evolution for $\mu_B=0$ of the propagator pole masses corresponding to those states, as given in the recent work \cite{Jung:2016yxl}. Note that the trend is similar  to the corresponding  lattice screening masses in Fig.\ref{fig:scrmas}. Actually, recent analysis confirm small deviations between screening and pole masses below $T_c$  within the PNJL model combined with lattice results  \cite{Ishii:2016dln}.

The case of the scalar channels is particularly important as far as CSR is concerned, since those include the quark condensate and the scalar susceptibility, directly reflecting  the response of the vacuum to symmetry breaking.   In that context, a relevant role  has been shown to be played by the thermal $f_0(500)$ pole generated in $\pi\pi$ scattering at finite temperature. Thus, from the thermal pole, one can define the scalar mass $M_S^2(T)=M_p^2(T)-\Gamma_p(T)^2/4=\re s_p(T)=\re \Sigma(s_p)$  with $\Sigma$  the self-energy of the $f_0(500)$ state. Therefore, we  expect such thermal mass to  inherit the  CSR properties expected from the $\sigma$ mode  \cite{Bochkarev:1995gi,Ferreres-Sole:2018djq}

\begin{figure}[h]
\resizebox{0.5\columnwidth}{!}{
\includegraphics{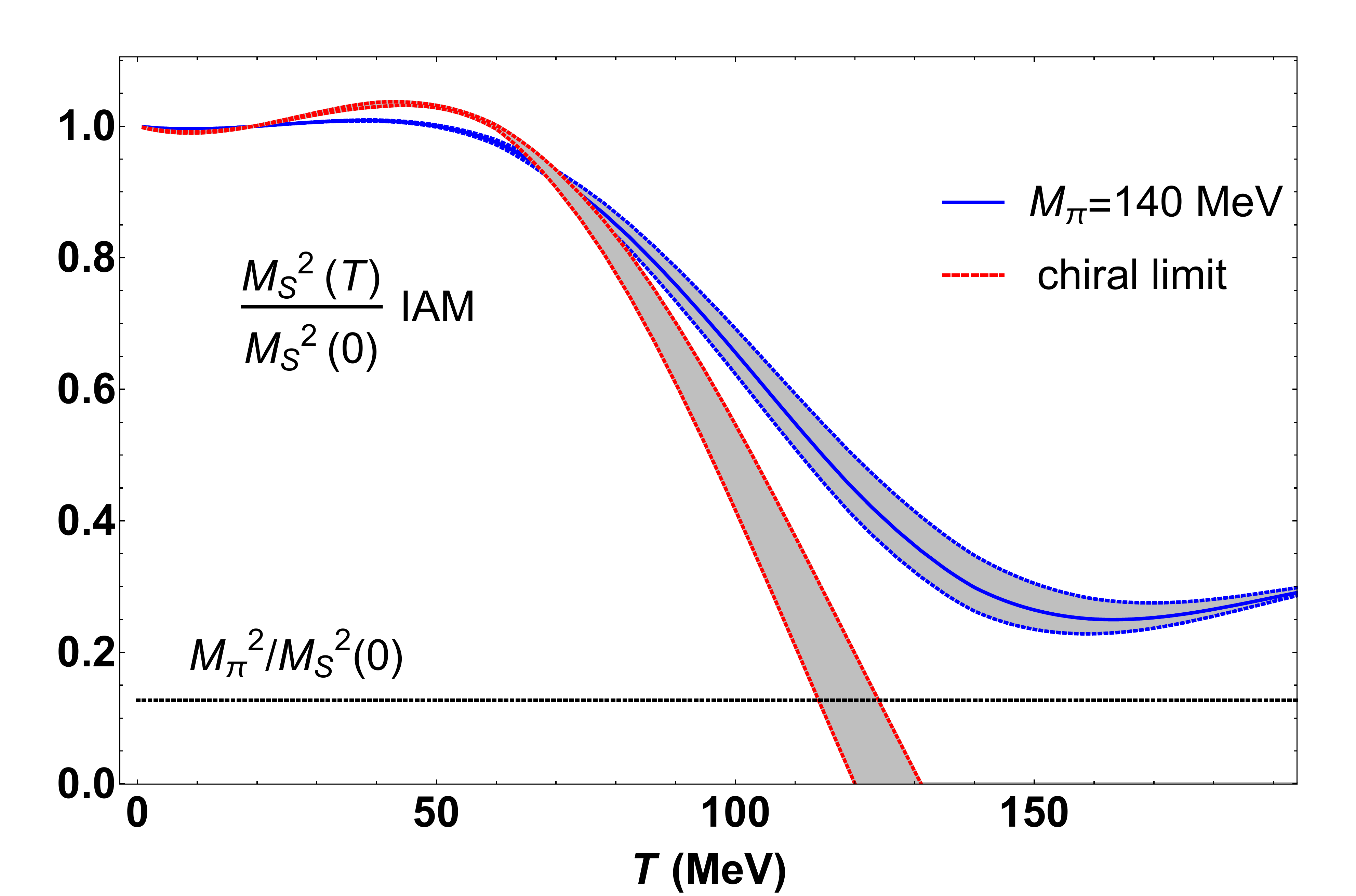}}
\resizebox{0.5\columnwidth}{!}{
\includegraphics{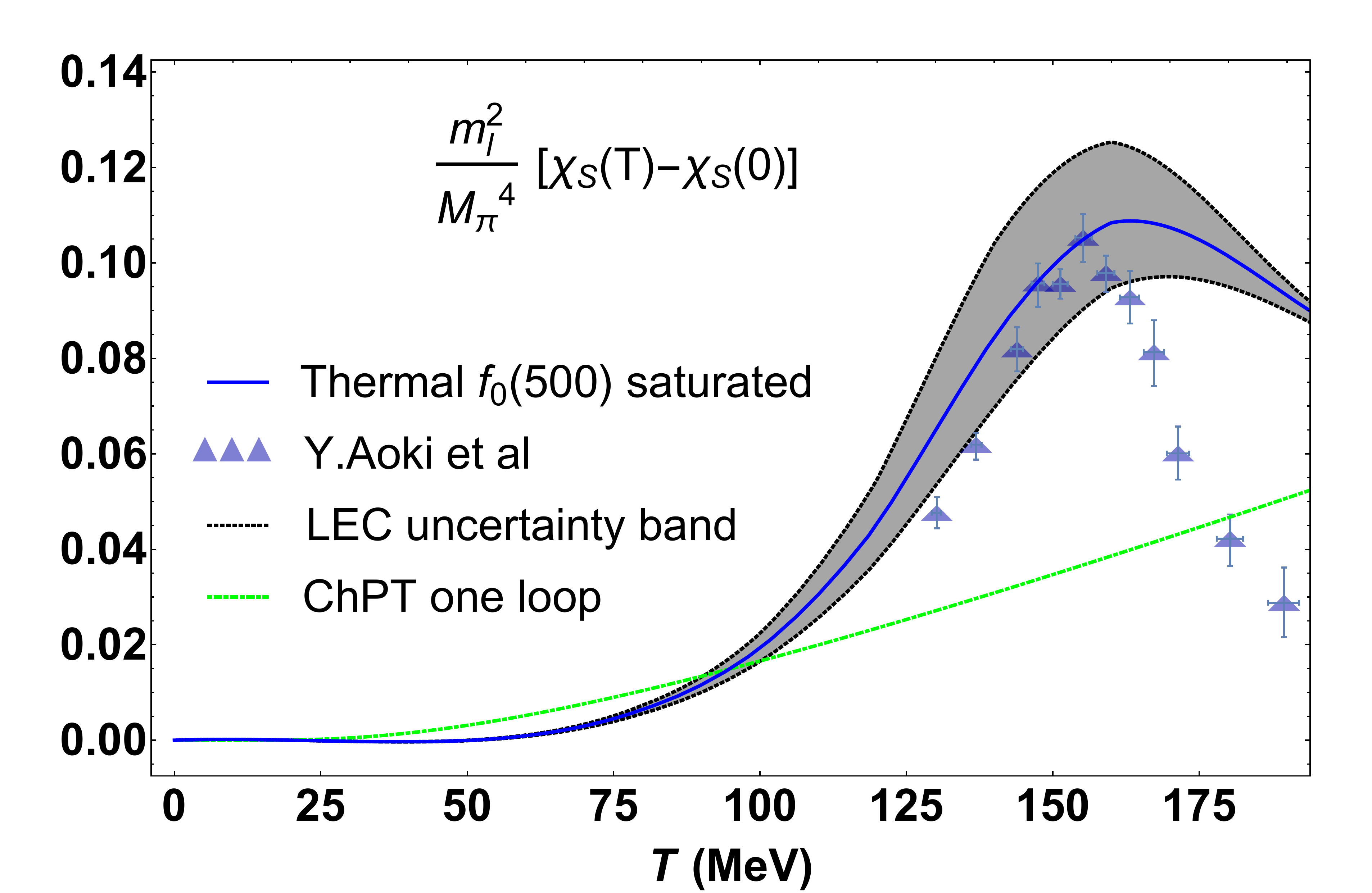}}
\caption{Unitarized scalar mass (left) and scalar susceptibilty (right) within the IAM thermal $f_0(500)$ pole approach.  The lattice points come  from \cite{Aoki:2009sc}. In the right panel the normalization of $\chi_S(0)$ is chosen to match the ChPT value in \cite{GomezNicola:2012uc}.} 
\label{fig:MsqandchiIAM}       
\end{figure}

In Fig.\ref{fig:MsqandchiIAM} we show the results for $M_S^2(T)$ from the so-called Inverse Amplitude Method (IAM) at finite temperature, for which 
the unitarized $I=J=0$ partial wave of the $\pi\pi$ scattering amplitude reads \cite{GomezNicola:2002tn,Dobado:2002xf} 
\begin{equation}
t_U(s;T)=\frac{t_2(s)^2}{t_2(s)-t_4(s;T)}
\end{equation}
with $s$ the Mandelstam variable and $t_2(s)+t_4(s;T)+\dots$ the ChPT expansion, where temperature enters through the one-loop diagrams in $t_4$. The above amplitude satisfies the thermal unitarity relation  $\im t_U(s;T)=\sigma_T(s) \vert t_U(s;T) \vert^2$ for $s\geq 4M_\pi^2$, where the thermal phase space is related to the $T=0$ one as  $\sigma_T(s;T)=\sigma(s;0)\left[1+2n(\sqrt{s}/2)\right]$ and $n(x)=\left(e^{x/T}-1\right)^{-1}$ is the Bose-Einstein distribution function. The latter has a neat interpretation in terms of two-particle states created and annihilated in the thermal bath \cite{GomezNicola:2002an}. 

As it can be seen in Fig.\ref{fig:MsqandchiIAM},  $M_S^2 (T)$ decreases in the physical pion mass case,  approaching the pion mass squared as a clear hint of partner degeneration. Recall that the pion mass is expected to depend softly on $T$ from ChPT-based analysis of the pion dispersion relation \cite{Schenk:1993ru}.  On the other hand, in the chiral limit, the temperature drop is more abrupt and tends to vanish at a temperature compatible with the lattice expectations discussed in section \ref{sec:signals}.

Furthermore, since the scalar susceptibility, as defined in \eqref{susdef},  corresponds to the scalar field propagator at vanishing momenta, it is expected to scale as the inverse squared scalar mass, which can be explicitly shown within the LSM context  \cite{Ferreres-Sole:2018djq}. Thus, saturating $\chi_S$ with the lightest thermal $f_0(500)$ state leads to the  unitarized or saturated susceptibility \cite{Nicola:2013vma,Ferreres-Sole:2018djq}
\begin{equation}
\frac{\chi_S^U (T)}{\chi_S^U (0)}=\frac{M_S^2(0)}{M_S^2(T)}
\label{unitsus}
\end{equation}
which, as can also be seen in Fig.\ref{fig:MsqandchiIAM}, is able to capture the essential behaviour of the lattice results, in particular the expected peak due to the crossover transition, within the uncertainties of the LEC involved \cite{Ferreres-Sole:2018djq}. The underlying assumptions are the smoothness of the $p$-dependence in the self-energy and of the $T$-dependence in  the  correlator normalization, the associated uncertainty lying within those of the LEC.  This approach describes lattice points in a very competitive way, compared for instance with the HRG case. Fits with both approaches are considered in \cite{Ferreres-Sole:2018djq}, the fits of the saturated case describing better the points around $T_c$ than the HRG, which captures  well the behaviour below the transition but gives rise to a monotonically increasing function, whereas the approach based on the thermal $f_0(500)$  reproduces clearly the expected peak, highlighting the importance of considering thermal interactions to understand the essential features of CSR, even though at those temperatures one would not expect such a description in terms of only the lightest d.o.f to work so accurately. In addition, the HRG is in conflict when trying to jointly fit  the quark condensate and the scalar susceptibility.  A similar scaling law as \eqref{unitsus} has been found to hold for the lattice screening masses in the $\pi$, $K$, $\bar s s$ and $\kappa$ channels, which combined with the use of WIs (see below) allow to explain quantitative and qualitatively the temperature dependence below and around $T_c$  \cite{Nicola:2013vma,Nicola:2016jlj,Nicola:2018vug,Nicola:2020iyl}. The latter  opens  additional interesting possibilities of connection with lattice  analyses,  which measure only screening masses, not expected to differ much from pole masses below the transition as commented above.

\section{The nature of the chiral  transition}
\label{sec:nature}

One of the  open challenges  regarding the QCD diagram is to properly understand  the nature or pattern of CSR.  In particular, both the universality class and the order of the transition  strongly depend   on the strength  of  $U(1)_A$ breaking at $T_c$.   Asymptotic  restoration of the $U(1)_A$ symmetry  with temperature, and therefore vanishing of the chiral anomaly,  is certainly a well-established theoretical possibility,  a possible mechanism being  the vanishing of the instanton density \cite{Gross:1980br}.  It must be pointed out  that in any case,  $U(1)_A$ symmetry breaking is by definition a short-distance or UV phenomenon, as it stems from its anomalous character. Therefore, there is no actual restoration transition at a given temperature, but a gradual fading away.  Precisely, the relevant question is to what extent  $U(1)_A$ is broken near the region of CSR.  If such breaking is sizable, a second order transition  in the $O(4)$ universality class is expected for $N_f=2$ in the chiral  limit, whereas partial or full $U(1)_A$ restoration at $T_c$ as the chiral limit is  approached  would lead to a $U(2)\times U(2)$ universality class corresponding to $O(4)\times U(1)_A$  restoration \cite{Pisarski:1983ms,Shuryak:1993ee,Cohen:1996ng,Pelissetto:2013hqa} and could even turn the transition into a first order one. In the  massive case, a residual $U(1)_A$ breaking is expected with respect to  massless quarks \cite{Lee:1996zy}.

\begin{figure}[h]
\hspace*{-0.4cm}
\resizebox{0.5\columnwidth}{!}{
\includegraphics{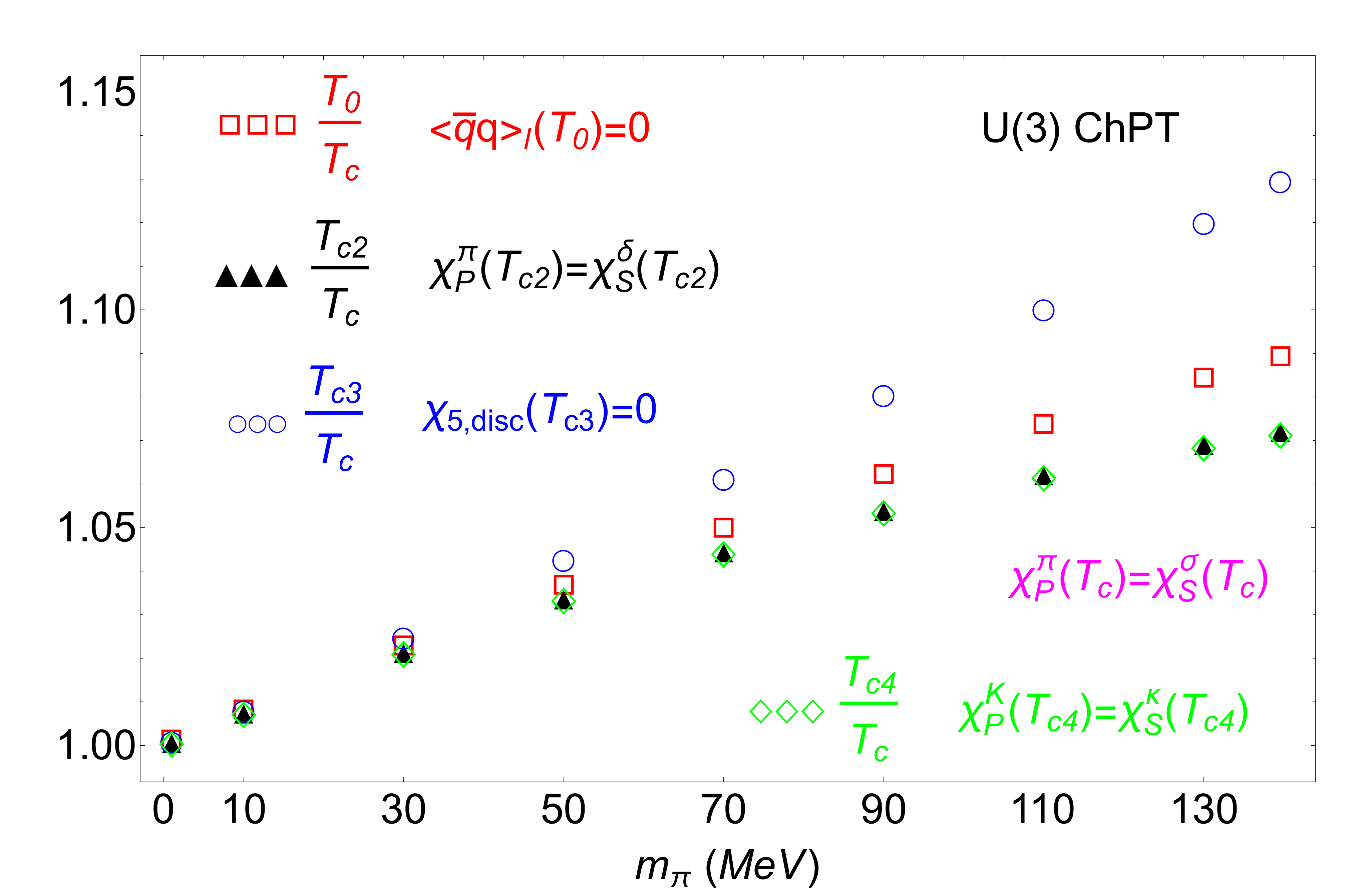}}
\hspace*{-0.5cm}
\resizebox{0.55\columnwidth}{!}{
\includegraphics{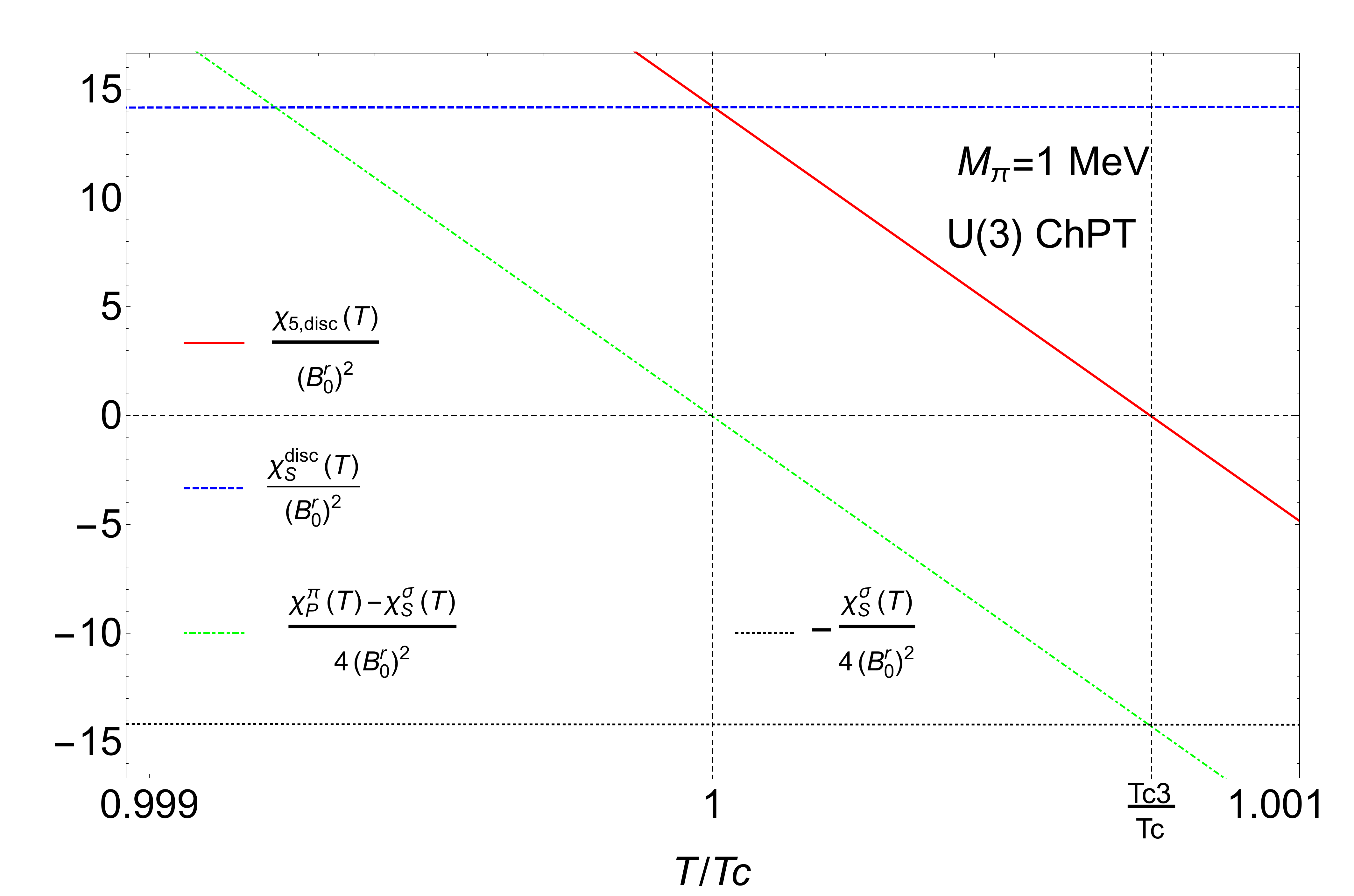}}
\caption{Results from the $U(3)$ ChPT analysis in \cite{Nicola:2018vug} for the evolution towards the chiral limit of pseudocritical temperatures corresponding to chiral and $O(4)\times U(1)_A$ degeneration (left) and for the dependence of the relevant susceptibilities near the chiral limit (right). } 
\label{fig:chiraltc}       
\end{figure}

This problem has been studied in several theoretical and lattice works  and there  is  still no full consensus about it. Most of the effort has been put on the fate of the chiral and $U(1)_A$ partners discussed in section \ref{sec:signals}. Thus, within LSM-like models,the interplay between chiral and $U(1)_A$ restoration and  partner degeneration for the $I=0,1$ nonstrange states in \eqref{nonets} under different restoring scenarios has been studied in \cite{Meggiolaro:2013swa,Heller:2015box} but results are not fully conclusive in  what  concerns $U(1)_A$ breaking at CSR. In addition, detailed studies of the different partners  have been recently conducted,  within the PNJL model combined with lattice results for the pole and screening masses  \cite{Costa:2008dp,Ishii:2015ira,Ishii:2016dln} and within the $U(3)$ ChPT scheme for the different susceptibilities involved \cite{Nicola:2018vug}. The latter  are in reasonable agreement with the lattice findings described in section \ref{sec:signals},  pointing to a sizable $U(1)_A$ breaking at $T_c$ for $N_f=2+1$ flavours in the physical limit, which should decrease or even disappear in the limit of two massless  flavours. In Fig.\ref{fig:chiraltc} we show the result in \cite{Nicola:2018vug} for the change with the pion mass of the  pseudocritical temperatures at which the different $O(4)$ and $O(4)\times U(1)_A$ partners degenerate,  clearly showing   their convergence in the chiral limit. It is also remarkable that $K-\kappa$ degeneration almost coincides with the $\pi-\delta$ one in this analysis,  therefore confirming  the $I=1/2$ sector as an alternative candidate to study $U(1)_A$ restoration. 

As commented in section \ref{sec:signals}, $N_f=2+1$ lattice simulations in the physical mass case observe a sizable gap between the degeneration of $O(4)$ and $U(1)_A$ partners, whereas for $N_f=2$ such gap is considerably reduced, being compatible with zero in the light chiral limit. In addition, a very recent $N_f=2+1$ analysis of the HotQCD group closer to the chiral limit \cite{Kaczmarek:2020sif} shows preliminary results indicating $U(1)_A$ breaking at $T_c$ based on the following relation:
\begin{equation}
\chi_{5,disc}(T)= \chi_S^{dis}(T)+ \frac{1}{4}\left[\chi_P^\pi(T)-\chi_S(T)\right]+\frac{1}{4}\left[\chi_S^\delta(T)-\chi_P^{\eta_l}(T)\right].
\label{chi5vschidisc}
\end{equation}
where $\chi_S=2\chi_S^{con}+4\chi_S^{dis}=\chi_S^{\delta}+4\chi_S^{dis}$ is the decomposition of the scalar susceptibility into quark connected and disconnected parts \cite{Nicola:2011gq,Nicola:2018vug}. Thus, in \cite{Kaczmarek:2020sif} it has been argued  that the peak in $\chi_S^{dis}$ coming from CSR would prevent the vanishing of $\chi_{5,disc}$ near $T_c$ since $\chi_P^\pi(T_c)=\chi_S(T_c)$ and $\chi_S^\delta$, $\chi_P^{\eta_l}$ are not meant to peak. 

However, the ChPT analysis in \cite{Nicola:2018vug} near $M_\pi\rightarrow 0^+$ shows that there would be no actual contradiction between $ \chi_{5,disc}$ peaking at $T_c$ and vanishing at $\chi_{5,disc}(T_{c3})=0$, where  $T_{c3}=T_c+\Od(M_\pi^2)$ around the chiral limit.  Actually,   $\chi_{5,disc} (T_c)$ and $ \chi_S^{dis}(T_c)$  do share the same leading  $\Od(T_c/M_\pi)$ coefficient in the expansion around $M_\pi\rightarrow 0^+$, consistently with both quantities peaking at $T=T_c$, since that coefficient is an indication of the critical behaviour within ChPT. However, at $T=T_{c3}$ the  $\Od(T_{c3}/M_\pi)$ coefficients that match are now those of $\chi_S^{dis}$ and $-\chi_S/4$,  rendering the right hand side in eq.\eqref{chi5vschidisc} regular (and actually vanishing) in the chiral limit at $T=T_{c3}$. In Fig.\ref{fig:chiraltc} we show the behaviour of those susceptibilities at a very small but nonzero pion mass, confirming the previous analysis. Therefore, near the chiral limit, $\chi_S^{dis}$ does  not really provide  information about $O(4)\times U(1)_A$ restoration, which should be rather analyzed through other observables like $\chi_{5,disc}$ itself, as we will discuss below. 

In any case, it must be borne in mind that the lattice analyses of  $U(1)_A$-related quantities carry a great deal of uncertainty, mostly in connection to the proper determination of  near-zero modes of the Dirac operator,  which dominate correlators involving the anomaly operator like the topological susceptibility (see below). For that reason, it is usually less noisy for lattice collaborations to measure differences of susceptibilities or screening masses connected by $U(1)_A$ rotations, such as $\chi_{ 5,disc}$.

A significant advance in this issue has come from the use of WIs  derived from the QCD generating functional and relating in a nontrivial and model-independent way susceptibilities  and quark condensates  \cite{Nicola:2013vma,Nicola:2016jlj,Azcoiti:2016zbi,GomezNicola:2017bhm,Nicola:2018vug,Nicola:2020iyl}. A particularly important WI among susceptibilities is the following:
\begin{equation}
\chi_P^{ls}(T)=-2\frac{m_l}{m_s} \chi_{5,disc}(T)=-\frac{2}{m_l m_s}\chi_{top}(T)
\label{WIls}
\end{equation}
where $\chi_P^{ls}=\int_T dx \langle \eta_l(x)\eta_s(0)\rangle$ , which is nonzero in the physical limit due  to $\eta-\eta'$ mixing, and 
the topological susceptibility is defined as $\chi_{top}=-\frac{1}{36}\int_T dx  \langle A(x) A(0) \rangle$ with $A(x)=\frac{3g^2}{32\pi^2}\epsilon_{\mu\nu\alpha\beta}G^{\mu\nu}_a  G^{\alpha\beta}_a$ the anomaly  operator and  $G^{\mu\nu}_a$ the gluon strength field tensor.  The connection between $\chi_{5,disc}$ and $\chi_{top}$ in eq.\eqref{WIls} has actually been  used in the lattice as a  check of the uncertainties involved  \cite{Buchoff:2013nra}.

The relevance of eq.\eqref{WIls} comes  from the observation that a $SU(2)_A$ rotation allows to transform the bilinear $\eta_l\rightarrow \delta$ and since  $\langle \delta \eta_l\rangle$  is a scalar-pseudoscalar correlator vanishing by parity, the conclusion is that $\chi_P^{ls}$ should vanish in the regime of exact CSR  and then, according to   \eqref{WIls}, both $\chi_{5,disc}$ and $\chi_{top}$ should vanish as well. We have seen in section \ref{sec:signals} that $\chi_{5,disc}$ measures $O(4)\times U(1)_A$ breaking. Hence,  for exact CSR, the $U(1)_A$ symmetry should also be restored (at least for the above partners) and the topological susceptibility should vanish. Those findings are in agreement with the $N_f=2$ lattice simulations in the chiral limit at the critical point, where CSR is meant to be exact. In addition, they are reproduced by the $U(3)$ ChPT analysis \cite{Nicola:2018vug}.  Moreover, the vanishing  of the  $ls$  correlator discussed above, implies ideal $\eta-\eta'$ mixing \cite{GomezNicola:2017bhm,Nicola:2018vug}, i.e, $\eta\sim \eta_l$, $\eta'\sim \sqrt{2}\eta_s$  and actually the mixing angle approaches the ideal value as temperature increases, both in  ChPT \cite{Nicola:2018vug} and in PNJL calculations \cite{Ishii:2016dln}. This is also consistent with the idea that the anomalous contribution to the $\eta'$ mass vanishes at ideal mixing  \cite{Guoetal}.

As we have just seen, the vanishing of the topological susceptibility with temperature is another sign of $U(1)_A$ restoration \cite{Azcoiti:2016zbi,GomezNicola:2017bhm,Nicola:2018vug} and  such vanishing is  indeed obtained in the lattice \cite{Bonati:2015vqz,Borsanyi:2016ksw,Lombardo:2020bvn} as well as theoretically, e.g, in ChPT both for $N_f=2$  \cite{diCortona:2015ldu}  and $N_f=2+1$ including the $\eta'$  \cite{Nicola:2019ohb},  and  in the PNJL model  \cite{Costa:2008dp}. The temperature dependence of this quantity has also important cosmological implications, given the direct relation between the topological susceptibility and the axion mass, namely $m_a^2=\chi_{top}/f_a^2$ with $m_a$ the axion mass, to leading order in $1/f_a$, the coupling of the axion field four-divergence to the $U(1)_A$ current \cite{diCortona:2015ldu}. Likewise, a reduction of the $\eta'$ mass is expected around the transition region, becoming  a ninth NGB in that regime with phenomenological implications \cite{Kapusta:1995ww}.  An indication of such reduction has been indirectly observed experimentally~\cite{Csorgo:2009pa}  from the analysis  of charged pion Bose-Einstein correlation data at low transverse momentum, that can be fitted with an enhanced $\eta'$ multiplicity consistent with a $M_{\eta'}$ reduction via thermal models. In addition, a decreasing behaviour for $M_{\eta'} (T)$ has been found in the lattice \cite{Kotov:2019dby}  and is also reproduced in  model calculations  \cite{Ishii:2016dln}. 

Additional insight  can be gained  from the $I=1/2$ sector regarding the role of strangeness and hence aiming to reconcile the $N_f=2$ and $N_f=2+1$ lattice results \cite{Nicola:2020wxy}. Thus, the following two WIs for the susceptibilities in that sector
\begin{eqnarray}
\chi_P^K(T)=-\frac{\condl (T)+2\conds (T)}{m_l + m_s}, \quad 
\chi_S^\kappa (T)=\frac{\condl (T)-2\conds (T)}{m_s-m_l},
  \label{wikappa} 
\end{eqnarray}
imply, on the one hand, that $\chi_P^K(T)$ should decrease monotonically, driven by the joint decrease of $\condl (T)$ and $\conds (T)$. On the other hand, the minus sign between the light and strange quark condensates in $\chi_S^\kappa (T)$ gives rise to  a peak for that susceptibility at a temperature above the transition at which the decreasing behaviour of $\conds(T)$ becomes stronger than the $\condl (T)$ one. The decreasing trend    after the peak leads $\chi_S^\kappa (T)$ towards the degeneration with $\chi_P^K(T)$ expected at $O(4)\times U(1)_A$ restoration, as commented before. 

\begin{figure}[h]
\hspace*{-0.35cm}
\resizebox{0.53\columnwidth}{!}{
\includegraphics{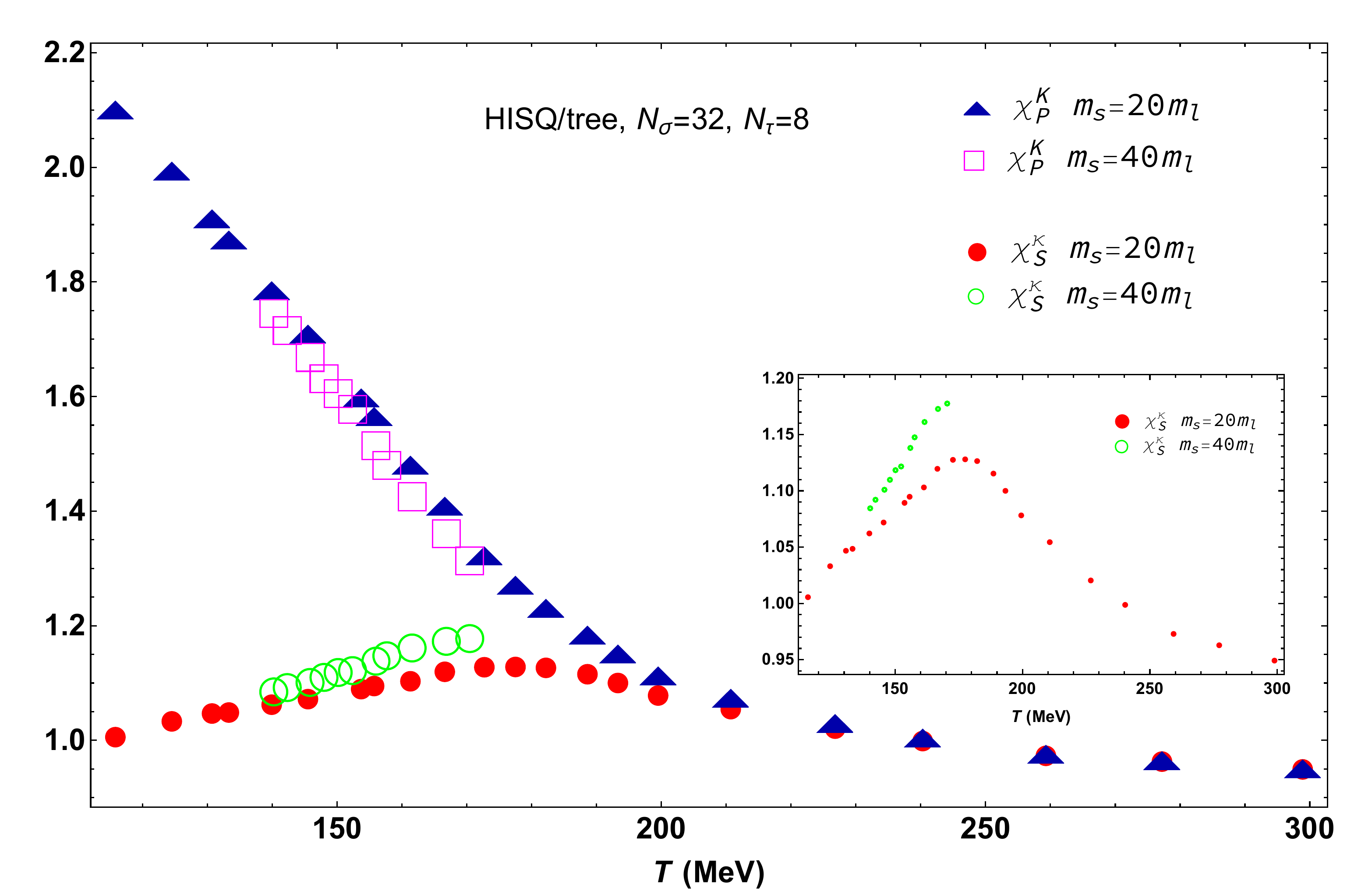}}
\hspace*{-0.35cm}
\resizebox{0.53\columnwidth}{!}{
\includegraphics{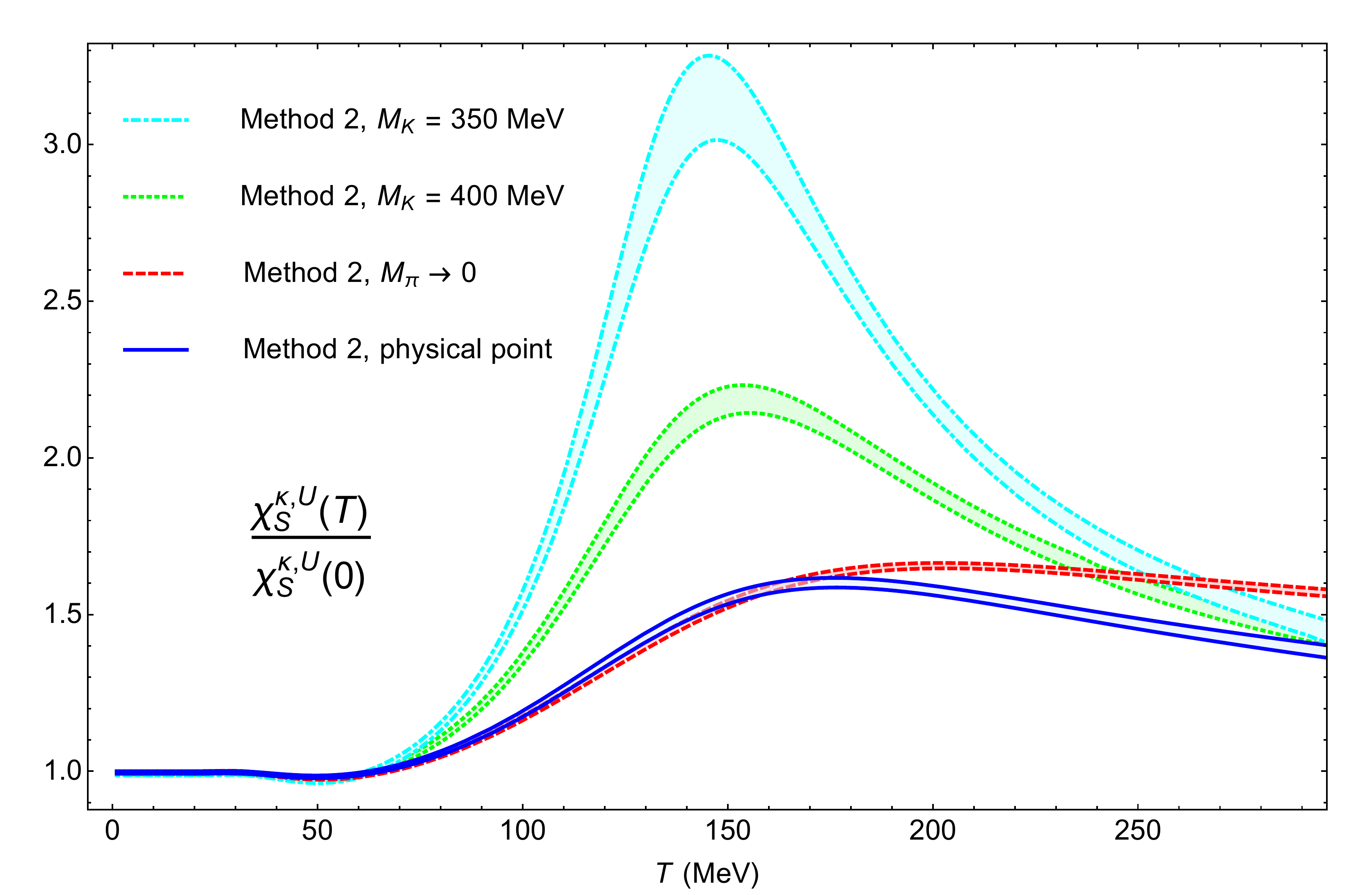}}
\caption{Results for susceptibilities in the $K-\kappa$ sector  \cite{Nicola:2020wxy}. Left panel: Susceptibilities extracted from the WIs in \eqref{wikappa} and the quark condensate lattice values in \cite{Bazavov:2011nk,Bazavov:2014pvz}. Right panel: Unitarized $\kappa$ susceptibility saturated  by the thermal  $K_0^* (700)$ pole in $\pi K$ scattering, for different pion and kaon masses.} 
\label{fig:Kkappa}       
\end{figure}

The above behaviour for the $K,\kappa$ susceptibilities can actually be observed for the corresponding quark condensate combinations in the lattice and explains in a quantitative way the role of strangeness in the gap between chiral and $U(1)_A$ restoration, ultimately controlled  by the $m_l/m_s$ ratio. Thus, near the light chiral limit, for vanishing $m_l/m_s$, the $\chi_S^\kappa (T)$ would flatten above the maximum as the $\conds$ decrease weakens and degeneration with $\chi_K$ becomes closer to $T_c$, thus recovering the $N_f=2$ chiral limit result for $O(4)\times U(1)_A$ restoration. For smaller strange quark mass, it would peak more abruptly,  resembling the $\chi_S$ behaviour in Fig.\ref{fig:MsqandchiIAM}.  In Fig.\ref{fig:Kkappa} we show  results  from \cite{Nicola:2020wxy} displaying such trends. We also show the result of a unitarized analysis performed in \cite{Nicola:2020wxy} where the $\kappa$ susceptibility is saturated by the thermal $K_0^*(700)$ pole generated in $\pi K$ scattering,  pretty much in the same spirit as in  the case of the scalar susceptibility discussed in section \ref{sec:theo}. The thermal evolution of the  $K_0^*(700)$ pole within the unitarized framework has actually been discussed recently in \cite{Gao:2019idb}.  Finally,  note that the minimum observed for the  screening masses   in this channel \cite{Cheng:2010fe,Bazavov:2019www,Ishii:2016dln,Nicola:2018vug}  is the counterpart of the maximum of the $\kappa$ scalar susceptibility  discussed here, thus revealing  an interesting global pattern for the scalar channels.   
 
\section{Conclusions}

The light quark sector plays an essential  role to properly understand  Chiral Symmetry Restoration. We have reviewed past and present knowledge about this key feature of the QCD phase diagram, mostly from lattice simulations and theoretical approaches. We have focused in some  hot topics that  nowadays remain challenging, like the interplay between chiral and $U(1)_A$ restoration and the role of the different partners involved, whose theoretical description has been investigated thoroughly over recent years through various techniques such as Effective Theories and Ward Identities. The Effective Theory framework captures the essential features of the chiral transition and supplemented with additional physical requirements such as unitarity allows to describe quite accurately the main observables involved. In fact, the role of thermal interactions ultimately leading  to modifications of the spectral properties of particles and resonances, turns out to be a crucial feature. The latter has allowed in particular to describe the peak of the scalar susceptibility around the transition saturating it  with the thermal $f_0(500)$ state generated in $\pi\pi$ scattering at finite temperature, accurately fitting  lattice points and even improving over the Hadron Resonance Gas description around the transition. Following the same ideas, the peak of the  $I=1/2$ scalar susceptibility expected from Ward Identities can be reproduced within the saturated approach with the $K_0^*(700)$  thermal pole in unitarized $\pi K$ scattering. 

The possibility of a significant $U(1)_A$ restoration around the chiral transition has concentrated most of recent efforts in this field, since it would have important theoretical and  phenomenological implications regarding e.g.  the universality class of the transition and the dependence with temperature of the topological susceptibility and of the $\eta'$ mass. Theoretical and lattice analyses point to a significant $U(1)_A$ restoration if only two light flavours are considered, which would become complete in the chiral limit. However, the situation in the physical case of $N_f=2+1$ case is not so clear,  both the chiral limit and the role of strangeness needing a better  understanding in that case. We have provided strong theoretical evidence based on Ward Identities and Effective Theories, which  on the one hand support full $U(1)_A$ restoration for exact chiral symmetry restoration, as for the case  of two massless flavours at $T_c$, and on the other hand  allow to quantitatively describe  the role of strangeness  through the influence of the strange quark condensate, connecting with well-measured lattice quantities. 

\section*{Acknowledgments}

Work partially supported by  research contracts FPA2016-75654-C2-2-P   (spanish ``Ministerio de Econom\'{\i}a y Competitividad") and PID2019-106080GB-C21(spanish ``Ministerio de Ciencia e Innovaci\'on"). This work has also received funding from the European Union Horizon 2020 research and innovation programme under grant agreement No 824093.    Useful discussion  and comments  from J.Ruiz de Elvira, A.Vioque, F.Karsch,  A.Lahiri,   M.P.Lombardo and J.A.Oller  are acknowledged. 

%

\end{document}